\documentclass[aps,prl,twocolumn,superscriptaddress,amsmath,amssymb,floatfix,longbibliography,nobibnotes]{revtex4-2}

\usepackage{newtxtext}\usepackage{newtxmath}
\usepackage{mathtools}
\usepackage{graphicx}
\usepackage{bm}
\usepackage{xcolor}
\usepackage{amsmath}
\usepackage{amssymb}
\usepackage{comment}
\usepackage{pict2e}

\usepackage{tikz}
\usetikzlibrary{arrows.meta}

\renewcommand\paragraph[1]{\emph{#1 ---}}

\usepackage[colorlinks=true,linkcolor=blue,citecolor=blue,urlcolor=blue]{hyperref}
\usepackage[capitalise]{cleveref}
\crefname{figure}{Fig.}{Figs.}  \Crefname{figure}{Figure}{Figures}
\crefname{equation}{Eq.}{Eqs.}  \Crefname{equation}{Equation}{Equations}
\crefname{appendix}{Appendix}{Appendices}  \Crefname{appendix}{Appendix}{Appendices}
\crefname{smsec}{Sec.}{Secs.}  \Crefname{smsec}{Section}{Sections}
\creflabelformat{equation}{(#2#1#3)}
\crefrangelabelformat{equation}{(#3#1#4)--(#5#2#6)}

\definecolor{tpgreen}{rgb}{0,0.5,0}

\newcommand{\hb}{\bar h}

\newcommand{\BB}{\mathcal{B}}   
\newcommand{\PP}{\mathcal{P}}   
\newcommand{\xig}{\xi_g}        
\newcommand{\dnn}{d_{\mathrm{nn}}}          

\usepackage{soul,xcolor}
\setstcolor{red}
\definecolor{lightblue}{rgb}{.90,.95,1}
\definecolor{greeny}{rgb}{0.9,1,0}
\sethlcolor{greeny}
\definecolor{grey}{gray}{0.55}

\newcommand{\papertitle}{Potts Order Preempts Emergence of 2-Form U(1) Spin Liquidity \\
in Three-Dimensional Spin Vorticity Models}

\newcommand{\authorMichael}{Michael D. Burke}
\newcommand{\authorMichel}{Michel J. P. Gingras}
\newcommand{\authorMohid}{Mohid U. Farooqi}
\newcommand{\authorKai}{Kristian Tyn Kai Chung}

\newcommand{\affiliationWaterloo}{Department of Physics and Astronomy,
University of Waterloo, Waterloo, Ontario, N2L 3G1, Canada}
\newcommand{\affiliationRice}{Department of Physics and Astronomy, Rice University, Houston, Texas 77005, USA}
\newcommand{\affiliationNano}{Waterloo Institute for Nanotechnology, University of Waterloo, Waterloo, Ontario, N2L 3G1, Canada}

\begin{document}

\author{\authorMichael}
\thanks{These authors contributed equally to this work.}
\affiliation{\affiliationWaterloo}
\author{\authorMohid}
\thanks{These authors contributed equally to this work.}
\affiliation{\affiliationWaterloo}
\author{\authorKai}
\affiliation{\affiliationRice}
\author{\authorMichel}
\affiliation{\affiliationWaterloo}
\affiliation{\affiliationNano}

\date{\today}

\title{
    \textcolor{black}{
    \papertitle
    }
}

\begin{abstract}
    Three-dimensional (3D) spin vorticity models (SVMs) have been proposed to realize an exotic 2-form U(1) classical spin liquid (CSL), characterized by a zero-curl constraint, string excitations, and deconfined fluctuating membranes.
    A previous numerical study [PRB {\bf 111}, 064417 (2025)] uncovered a puzzle in the pyrochlore lattice SVM: despite exhibiting  pinch points characteristic of a U(1) spin liquid, it also develops weak all-in--all-out (AIAO) order.
    Here, we show that the emergence of the 2-form CSL in 3D SVMs is preempted by an intrinsic instability towards entropy-driven height locking.
    We establish this by resolving the ground state constraints using a height mapping, which we further relate to a 3-state Potts antiferromagnet.
    This identifies the pyrochlore's weak AIAO order as permutationally symmetric sublattice (PSS) order, albeit with an exceptionally weak locking strength.
    We extend the SVM to lattices of corner-sharing octahedra and cubes, where we identify the alternative broken-sublattice-symmetry (BSS) order, a hidden order without magnetic Bragg peaks coexisting with collective paramagnetism. 
    All three models undergo a 3D Ising transition consistent with a symmetry-allowed locking potential in the coarse-grained height theory. The spin liquid persists over intermediate length and temperature scales, but a critical Coulomb ground state ensemble fails to materialize, its correlations cut off by the finite locking length.
    These findings illustrate how microscopic constraints can determine the ultimate stability of emergent gauge theories in frustrated spin systems.
\end{abstract}

\maketitle

\begingroup
\renewcommand{\thefootnote}{*}
\footnotetext{These authors contributed equally to this work.}
\endgroup

Classical spin liquids are correlated paramagnets~\cite{villainInsulatingSpinGlasses1979} in which local energetic constraints prevent conventional magnetic order down to the lowest temperatures, while organizing the spins into a strongly fluctuating state with an emergent gauge structure and critical correlations~\cite{MoessnerChalker,IsakovGregor,henleyCoulombPhaseFrustrated2010,yanClassificationClassicalSpin2024,yanClassificationClassicalSpin2024a,lozano-gomezAtlasClassicalPyrochlore2024,henleyPowerlawSpinCorrelations2005}.
Among the best-studied examples is spin ice, realized by Ising spins on the pyrochlore lattice of corner-sharing tetrahedra, which has proved a remarkably fertile setting for the study of unconventional phenomena in frustrated magnetism~\cite{HarrisSpinIce,bramwellSpinIceState2001,CastelnovoAnnRevCMP}. 
Its extensively degenerate ground-state manifold consists of spin configurations satisfying the local ice rule---a zero-divergence constraint on the magnetic flux field---giving rise to a 
U(1) gauge theory in which the ground state can be viewed as a condensate of closed strings, while excitations take the form of point-like magnetic monopoles at the ends of open strings~\cite{henleyCoulombPhaseFrustrated2010,Castelnovo2008,gingrasObservingMonopolesMagnetic2009,morrisDiracStringsMagnetic2009,moessnerSpinIceCoulomb2021,niggemannClassicalFractonSpin2026,plackeIsingFractonSpin2024}.
The introduction of transverse quantum fluctuations into this classical setting has motivated extensive theoretical and experimental research on quantum spin ice, where a quantum U(1) spin liquid with gapless photon excitations can arise~\cite{HermelePhoton,MolavianQSI,SavaryCoulomb,BentonLight,GingrasQSIRPP,Gao2025}.
Building on this gauge-theoretic picture, considerable progress has been made in understanding higher-rank tensor spin liquids, both classical and quantum, predominantly described by symmetric tensor gauge fields~\cite{bentonSpinliquidPinchlineSingularities2016,PretkoFractons,YanRank2,lozanogomez2023arxiv,sternFractonSpinLiquid2026a}. A particularly interesting case are 2-form U(1) spin liquids, whose gauge field is an antisymmetric tensor: the condensate of closed strings is promoted to a condensate of closed membranes, and point-like monopoles give way to string-like excitations~\cite{KalbRamond,henneauxPFormElectrodynamics1986,LozanoKalbRamond}.

The spin vorticity model (SVM)~\cite{chung2formU1Spin2025,RougemailleAmpere} is a classical Ising spin model whose interactions impose a vanishing curl constraint on ground states, and offers a candidate realization of a classical 2-form U(1) spin liquid.
For the anticipated gauge-theoretic spin liquid to be thermodynamically stable, the system must remain in a deconfined phase, characterized by algebraically decaying correlations and associated pinch-point singularities in the equal-time spin structure factor $S(\bm q)$. 
A previous Monte Carlo study of the pyrochlore SVM~\cite{chung2formU1Spin2025} reported encouraging signatures of spin-liquid behavior: pinch-point-like features qualitatively consistent with predictions from the self-consistent Gaussian approximation (SCGA), together with an extensive residual entropy.
Surprisingly, the same simulations also revealed a sharp finite-temperature anomaly accompanied by the onset of all-in--all-out (AIAO) magnetic long-range order, albeit with an order parameter that remains far below its saturation value down to zero temperature.

These findings suggest an unusual coexistence of spin-liquid-like correlations and weak magnetic long-range order, raising several fundamental questions.
What is the nature of the transition at the specific heat peak, and does it involve the confinement of string-like defects?
What drives the weak AIAO order, and does it cause the transition or merely accompany it?
Most importantly, can magnetic order coexist with the algebraic correlations expected of a 2-form U(1) spin liquid, or does it suppress their pinch-point singularities in the thermodynamic limit?
We address these questions by studying the SVM on three different three-dimensional lattices of corner-sharing clusters: pyrochlore (tetrahedra), octochlore (octahedra), and cubichlore (cubes).
Using an integer-valued height representation, we establish a mapping between the SVM ground-state manifolds and those of the three-state Potts antiferromagnet on their respective parent lattices~\cite{salasThreeStateSquareLatticePotts1998,koteckyEntropyDrivenPhaseTransition2014,lapinskasTwoTypesOrdering1998,banavarOrderingPhaseTransitions1980,wangThreestateAntiferromagneticPotts1990,gottlobThreestateAntiferromagneticPotts1994,kurbanovaCriticalPropertiesAntiferromagnetic2025}.
We show that a relevant, entropy-driven height-locking potential induces lattice symmetry breaking at nonzero temperature.
Within this description, the weak AIAO order of the pyrochlore SVM corresponds to the permutationally symmetric sublattice (PSS) phase of the Potts antiferromagnet on the diamond lattice~\cite{lapinskasTwoTypesOrdering1998,oshikawaOrderedPhaseScaling2000}. 
On the octochlore and cubichlore, with simple-cubic and body-centered-cubic parent lattices, respectively, the mapping instead points to the broken-sublattice-symmetry (BSS) phase~\cite{banavarOrderingPhaseTransitions1980,wangThreestateAntiferromagneticPotts1990,kurbanovaCriticalPropertiesAntiferromagnetic2025}, which we confirm numerically.
Large-scale Monte Carlo simulations further provide evidence that height locking renders spin correlations finite-ranged, cutting off development of pinch-point singularities at low temperatures.
Our work exposes an intrinsic instability of SVMs: the emergent gauge theory survives as an intermediate-scale description but is rendered massive at long distances.

\begin{figure*}[t]
  \setlength{\unitlength}{1pt}
  \begin{picture}(508,204)
    \put(0,0){\includegraphics{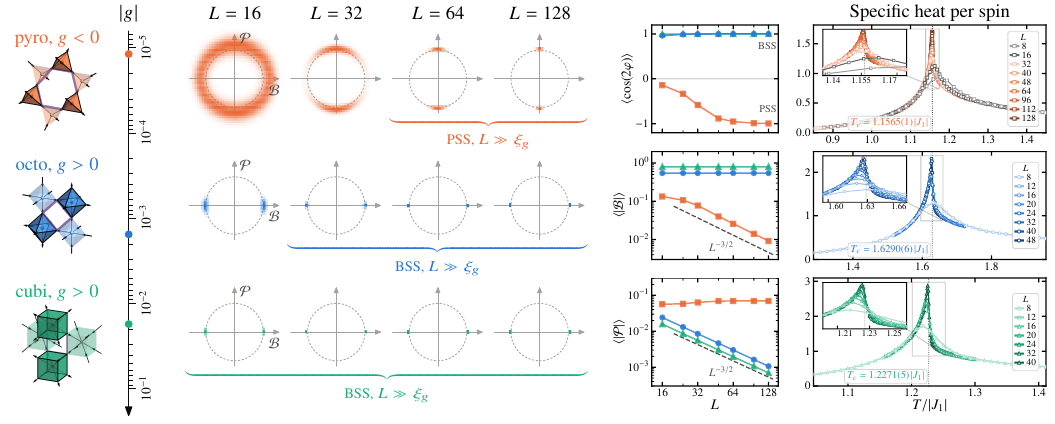}}
    \put(0,202){(a)}\put(86,202){(b)}\put(288,202){(c)}\put(288,133){(d)}\put(288,72){(e)}\put(378,202){(f)}
  \end{picture}
  \caption{
  Lattices, ground-state ensembles and specific heat.
  (a) Minimal plaquettes (purple) of each lattice/parent: pyrochlore/diamond (orange), octochlore/simple cubic (blue), cubichlore/bcc (green), ordered by increasing $|g|$ obtained from $T=0$ height fits and denoted by dots on the log-scale axis.
  (b) Joint histograms of $(\BB,\PP)$ at $T=0$ for $L=16$--$128$, scaled such that $L=128$ peaks lie on the unit circle. 
  Opacity shows the number of samples in each bin (linear scale), symmetrized over $\BB\to-\BB$ and $\PP\to-\PP$.
  (c) $\langle\cos(2\varphi)\rangle$ against $L$: $+1$ (BSS) on the octochlore and cubichlore lattices, $-1$ (PSS) on the pyrochlore lattice.
  (d) $\langle|\BB|\rangle$ and (e) $\langle|\PP|\rangle$ against $L$.
  (f) Specific heat per spin against $T/|J_1|$ for $L=8$--$128$. Inset: zoom-in of the peak.
  }
  \label{fig:T0}
\end{figure*}

\paragraph{Models} 
We consider the pyrochlore, octochlore~\cite{henleyCoulombPhaseFrustrated2010,sternFractonSpinLiquid2026a,szaboFragmentedSpinIce2022,sklanNonplanarGroundStates2013,bentonTopologicalRouteNew2021,niggemannQuantumEffectsUnconventional2023}, and cubichlore lattices of corner-sharing tetrahedra, octahedra and cubes, respectively.
These are the medial or line graph lattices of bipartite parent lattices~\cite{henleyCoulombPhaseFrustrated2010} (diamond, simple cubic and body-centered cubic), with coordination $z=4,6,8$ and nearest-neighbor distance $\dnn$.
The clusters reside on the parent vertices and the spins on the parent bonds.
The SVM Hamiltonian is
\begin{equation}
  H = \frac{J}{2}\sum_p \omega_p^2, 
  \qquad 
  \omega_p \equiv \mathrm{curl}(\sigma)_p
  = \sum_{i\in p} \epsilon_{p,i}\, \sigma_i  ,
  \label{eq:H}
\end{equation}
where $J>0$ sets the energy scale, $p$ indexes the plaquettes of the parent lattice, $\cramped{\sigma_i = \pm 1}$ are Ising spins located at sites $i$, with quantization axes along the parent bonds [arrows in \cref{fig:T0}(a)], and $\omega_p$ is the \emph{spin vorticity}, or lattice curl, where $\cramped{\epsilon_{p,i}=\pm1}$ alternates around $p$.
Expanding \cref{eq:H} yields Ising models with couplings up to third neighbors on the pyrochlore and second neighbors on the octochlore and cubichlore lattices~\cite{SM};
we quote temperatures in units of $|J_1|$.
On each lattice the interaction matrix has two degenerate flat bands at its minimum, signaling a macroscopically degenerate ground-state manifold~\cite{SM}. 
On these grounds, it was argued in Ref.~\cite{chung2formU1Spin2025} that the SVM on the pyrochlore lattice realizes a classical spin liquid, the excitations of which are one-dimensional vortex strings rather than particles and whose gauge degrees of freedom are two-dimensional membranes.

\begin{figure}[t]
  \setlength{\unitlength}{1pt}
  \begin{picture}(246,114)
    \put(0,0){\includegraphics{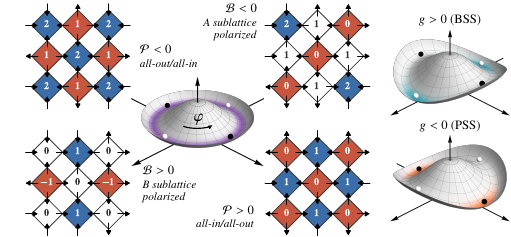}}
    \put(0,108){(a)}\put(178,108){(b)}
  \end{picture}
  \caption{(a) Free energy of the height angle $\varphi=\pi\hb$ over the order-parameter plane $(\BB,\PP)$ at $g=0$: the minimum is a degenerate ring (purple) on which $\varphi$ is free. Representative spin configurations illustrate the heights $h_v$ [\cref{eq:height}]---even (odd) on the A (B) sublattice for the BSS and PSS orders. 
  Red (blue) clusters are polarized all-out (all-in), and white clusters are not fully polarized. 
  PSS order ($\varphi=\pm\pi/2$, black markers) corresponds to AIAO order; for $\PP>0$ the A clusters are all-out. 
  BSS order ($\varphi=0,\pi$, white markers) has one sublattice polarized but fluctuating---A (B) for $\BB$ negative (positive). (b) The $\cos(2\varphi)$ locking anisotropy [\cref{eq:free_energy}] corrugates the ring: positive (negative) $g$ selects BSS (PSS) order.}
  \label{fig:BSS_PSS}
\end{figure}

\paragraph{Height field and Potts mapping}
Focusing on the zero-curl constraint, $\omega_p=0$, we see that ground state spin configurations can be expressed as the lattice gradient of an integer-valued  height field $h_v$ on the vertices $v$ of the bipartite parent lattice, with A and B sublattices:
\begin{equation}
  \sigma_{\langle vv'\rangle} = h_{v'} - h_v , \qquad v\in A,\ v'\in B .
  \label{eq:height}
\end{equation}
We fix $\cramped{h_{v_0}=0}$ at a reference vertex $\cramped{v_0 \in A}$, then construct $h_v$ by summing $\pm\sigma$ along any path from $v_0$ to $v$. 
Path-independence follows because the sum around any contractible loop equals the total $\omega_p$ of the plaquettes it encloses, which vanishes~\cite{SM}.
Since every step changes $h$ by $\pm 1$, $h$ is even (odd) on the A (B) sublattice.
Each spin configuration therefore fixes $h$ up to a uniform even-integer shift to preserve this parity, giving a compact height field of period 2.
In this representation, the excitations are the vortices of~$h$, while its level sets describe the closed-membrane gauge degrees of freedom.

The height mapping provides insight into the structure of the ground state manifold. 
On all three lattices, $\cramped{c_v = h_v\bmod 3}$ defines a proper 3-coloring of the parent lattice; this establishes a 3-to-1 correspondence between ground states of the 3-state Potts antiferromagnet and the SVM (with an additional constraint for the pyrochlore case) [\cref{app:potts}].
The Potts model is known to exhibit either broken sublattice symmetry (BSS) order or permutationally symmetric sublattice (PSS) order on these lattices~\cite{banavarOrderingPhaseTransitions1980,koteckyLongrangeOrderAntiferromagnetic1985,wangAntiferromagneticPottsModels1989a,wangThreestateAntiferromagneticPotts1990,gottlobThreestateAntiferromagneticPotts1994,rahmanIntermediatetemperatureOrderingThreestate1998,kurbanovaCriticalPropertiesAntiferromagnetic2025,lapinskasTwoTypesOrdering1998,junFerrimagneticPhasesBlumeEmeryGriffiths2000}.
The ground state correspondence indicates that the SVM must also harbor order at $\cramped{T=0}$.

\paragraph{Height locking} To understand the mechanism behind the SVM's order and its implications for spin liquidity, we develop a continuum theory of the height field.
Letting $\bar{h}(\bm{r})$ be the average of the heights in a block around position $\bm{r}$~\cite{kondevFourcoloringModelSquare1995}, we obtain a slowly varying compact field angle $\varphi(\bm{r}) = \pi \bar{h}(\bm{r})$, defined modulo $2\pi$ by the height-shift redundancy.
Sublattice exchange changes the parity of $h$ and so sends $\varphi\to\varphi+\pi$, while time reversal acts as $\varphi\to-\varphi$. 
Together, these symmetries constrain the leading free energy density to
\begin{equation}
  \mathcal{F}/T = \frac{K}{2\pi^2}(\nabla\varphi)^2 - g\cos(2\varphi) ,
  \label{eq:free_energy}
\end{equation}
where $K$ is the entropic stiffness, set by the number of microstates compatible with a given coarse-grained height gradient.
In three dimensions the anisotropy term is relevant at the Gaussian fixed point, causing the height field to lock: $\cramped{\varphi\in\{0,\pi\}}$ for $\cramped{g>0}$, or $\cramped{\varphi\in\{\pm\pi/2\}}$ for $\cramped{g<0}$ [\cref{fig:BSS_PSS}].
Expanding about these minima gives the mass of the height fluctuations and a corresponding locking length scale, $\cramped{\xig=(K/4\pi^2|g|)^{1/2}}$, below which the system exhibits approximately algebraic spin liquid correlations, crossing over to locked behavior beyond it.

The two locking patterns correspond to distinct forms of microscopic order, captured by the real and imaginary components of the height order parameter, $\exp(i\varphi)$. 
Their corresponding microscopic spin observables are
\begin{equation}
    \BB = 
    \frac{1}{L^3}
    \bigg(
    \sum_{v\in B}m_v^2 - \sum_{v\in A}m_v^2\bigg),
    \qquad 
    \PP=\frac{1}{L^3}\sum_{v \in A}m_v ,
    \label{eq:BP}
\end{equation}
where
$\cramped{m_v = z^{-1} \sum_{i \ni v} \sigma_i}$
is the net AIAO moment around vertex $v$, assuming a system of $L^3$ primitive unit cells.
The sublattice imbalance $\cramped{\BB\propto\cos\varphi}$ is even under time reversal, while $\cramped{\PP\propto\sin\varphi}$ (the AIAO order parameter) is odd.
The two types of order are best illustrated with characteristic configurations [\cref{fig:BSS_PSS}]. 
For $\cramped{g<0}$, half-odd-integer locking is exemplified by $\cramped{h_A=0}$ and $\cramped{h_B=1}$ so  that $\cramped{\bar{h}=1/2}$, yielding uniformly polarized spins and PSS (equivalently, AIAO) order with $\cramped{\BB=0}$ and $\cramped{\PP\neq0}$.
For $\cramped{g>0}$, the coarse-grained height locks to integer values, as exemplified by $\cramped{h_A=0}$ and $\cramped{h_B=\pm1}$ with random signs so that $\cramped{\bar{h} = 0}$: one sublattice is locked while the other has locally saturated but fluctuating moments, realizing BSS order with $\cramped{\BB\neq0}$ and $\cramped{\PP=0}$. 
See \cite{SM} for further details.

\paragraph{Ground-state ensembles} We now test the height-field description against Monte Carlo ground-state ensembles of the SVM, extracting the sign and magnitude of the locking potential and examining its consequences for ordering and spin-liquid correlations. We measure the spin order parameters $\BB$ and $\PP$ [\cref{eq:BP}] directly, and independently reconstruct the integer heights $h_v$ of each sampled configuration. From these we obtain the global height angle $\cramped{\varphi=\pi\hb}$, where $\hb$ is the mean height of the entire system.
We sample ground states directly via Monte Carlo simulation for linear system sizes $L$ up to 128 using energy-conserving star moves~\cite{chung2formU1Spin2025}---reversing all spins in a fully-polarized cluster, shifting $h_v$ by~$\pm2$~\cite{SM}.

\begin{figure*}[t]
  \centering
  \setlength{\unitlength}{0.001968\textwidth}
  \begin{picture}(508,150.8)
    \put(0,0){\includegraphics{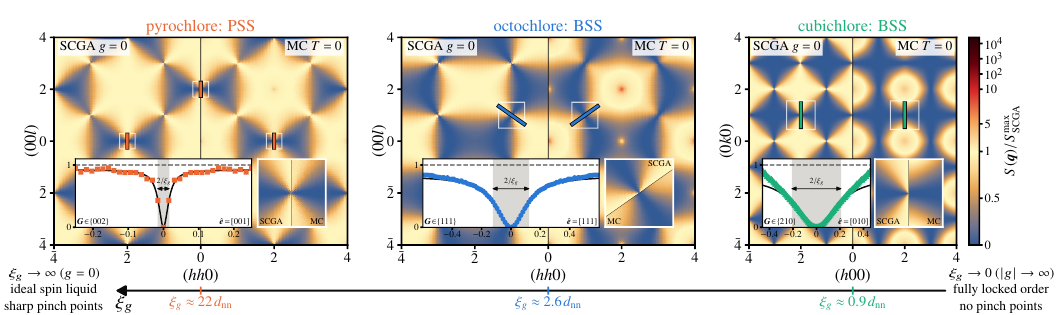}}
    \put(4,138.8){(a)}\put(170,138.8){(b)}\put(337,138.8){(c)}
  \end{picture}
  \caption{
  Pinch-point smearing by the height-locking. 
  Spin structure factor at $T=0$ in the
  $(hhl)$ plane of the (a) pyrochlore and (b) octochlore and the $(hk0)$ plane of the (c) cubichlore SVM. 
  The arrow orders the three lattices by the locking length $\xig$, from the ideal spin liquid ($\xig\to\infty$, $g=0$: sharp pinch points) to full locking ($\xig\to0$: no pinch points).
  Left halves: SCGA
  at $g=0$~\cite{SM}; right halves: Monte Carlo at $T=0$ ($L=128$).
  The color scale is linear from $0$ to $1$ (the SCGA maximum) and logarithmic beyond.
  Left insets: line cuts $\bm q=\bm G+\delta q\,\hat{\bm e}$ through a pinch point $\bm G$ along the pinch-point direction $\hat{\bm e}$ (both given in each inset; colored bars on the maps).
  Dashed line: SCGA; points: MC; solid line: fits to $A\,\delta q^2/(\delta q^2+\xig^{-2})$ with $\xig$ fixed by the $T=0$ height-field fits and only $A$ adjusted; shaded region: $|\delta q|<1/\xig$.
  Right insets: zoom on the pinch points, juxtaposing SCGA (sharp) and MC (smeared) pinches. 
  }
  \label{fig:pinch}
\end{figure*}

The ground-state ensembles reveal a striking contrast between the three lattices: the octochlore and cubichlore favor BSS order, while the pyrochlore selects PSS order. 
This is captured by the height observable $\langle\cos(2\varphi)\rangle$ [\cref{fig:T0}(c)], which approaches $+1$ for BSS and $-1$ for PSS, directly identifying the sign of the locking potential $-g\cos(2\varphi)$. 
On the octochlore and cubichlore lattices ($\cramped{g>0}$), it is already close to $+1$ for the smallest sizes, whereas on the pyrochlore ($\cramped{g<0}$) it approaches $-1$ remarkably slowly, reaching $-0.993$ only at $\cramped{L=128}$. 
The joint histograms of the spin observables $\cramped{(\BB,\PP)}$ [\cref{fig:T0}(b)] corroborate this: the pyrochlore evolves from a nearly circular distribution at $\cramped{L=16}$ into two distinct PSS peaks at $\cramped{L=128}$, while the octochlore and cubichlore exhibit two BSS peaks even at the smallest sizes, with sharper peaks on the cubichlore. 
Consistently, the disordered component of the order parameter decays as $\cramped{L^{-3/2}}$ while the ordered component saturates [\cref{fig:T0}(d,e)], reaching $\cramped{|\PP|=0.070}$ on the pyrochlore and $\cramped{|\BB|=0.545}$ (octochlore) and $0.792$ (cubichlore) at $L=128$.

To determine the strength $\vert g \vert$ of the  height locking potential, we turn to the height-height correlations. 
At quadratic order about a locked minimum, \cref{eq:free_energy} reduces to a massive Gaussian theory with ${\mathcal{F}}/T \simeq(K/2)(\nabla h)^2+2\pi^2|g|\,\delta h^2$, with $\delta h$ the deviation of the coarse-grained height from its locked value, giving the momentum space height correlator $\cramped{\langle|h({\bm q})|^2\rangle\propto(K|\bm q|^2+4\pi^2|g|)^{-1}}$ at small $|\bm q|$.
Fitting this to the $\cramped{T=0}$ ensembles~\cite{SM}, we obtain the stiffnesses $\cramped{K=0.233}$, $0.409$, $0.526$ and locking strengths $\cramped{|g|=1.2\times10^{-5}}$, $1.5\times10^{-3}$, $1.7\times10^{-2}$ for the pyrochlore, octochlore and cubichlore, respectively, giving the locking lengthscales $\cramped{\xig\approx22\,\dnn}$, $2.6\,\dnn$ and $0.9\,\dnn$, in units of the nearest-neighbor distance $\dnn$, or about $14$, $2$ and $1$ primitive unit cells, the unit of $L$.

The locking on the pyrochlore lattice is remarkably weak, two orders of magnitude weaker than the other lattices, resulting in a large $\xi_g$.
This produces severe finite-size effects, visible in the scaling of the order parameters~[\cref{fig:T0}(b-e)]: $\langle \cos(2\varphi)\rangle$ saturates only for system sizes larger than $\cramped{L \sim 64}$.
The most striking feature of the pyrochlore's PSS order is how little of it the spins experience: at $\cramped{L=128}$ the height angle is essentially fully locked, $\langle\cos(2\varphi)\rangle=-0.993$, yet the AIAO moment saturates at only $\cramped{|\PP|=0.070}$, $7\%$ of its maximum value~\cite{chung2formU1Spin2025}.
This suppression is entropic: the order lives in the heights, not in the spins---locking the mean height leaves significant room for spin fluctuations, so they carry only a small coherent bias.
A measurement of the spins alone would indicate a barely ordered magnet, whereas in the height variables it is fully ordered; the smallness of the AIAO moment is a property of the ordered state itself.

Both the locking strength and the selected order correlate with coordination number, as do the PSS and BSS phases of the $q=3$ Potts antiferromagnet~\cite{lapinskasTwoTypesOrdering1998}.
The microscopic mechanism that fixes the sign of~$g$, however, remains unclear. An analogous situation arises in the Landau theory of spin-Peierls order on the pyrochlore lattice, which admits a similar mapping onto a Potts antiferromagnet on the diamond lattice~\cite{tchernyshyovSpinPeierlsPhasesPyrochlore2002}. 
There, too, the sign of the sixfold anisotropy that selects between BSS-like and PSS-like distortions is not fixed by symmetry.

\paragraph{Pinch-point smearing}
Upon cooling, the suppression of thermally excited gauge charges allows the correlations of the 2-form Coulomb spin liquid to develop, but height locking ultimately preempts a critical ground-state ensemble. 
Two length scales govern this competition: a thermal length $\xi_{\mathrm{th}}$ set by the density of gauge charge strings, which grows upon cooling~\cite{IsakovGregor,henleyCoulombPhaseFrustrated2010}, and the nearly temperature-independent locking length $\xig$. 
In three dimensions, symmetry-allowed periodic potentials generically lock the scalar height field~\cite{polyakovQuarkConfinementTopology1977a,gopfertProofConfinementStatic1982}, imparting an entropic tension to the membranes and rendering the gauge fluctuations massive beyond $\xig$. 
The pinch points therefore sharpen upon cooling as $\xi_{\mathrm{th}}$ grows, but cease to sharpen once $\cramped{\xi_{\mathrm{th}}\sim\xig}$ and retain a finite width to $T=0$. 
The large locking length on the pyrochlore permits an extended finite-temperature spin-liquid regime, whereas the cubichlore exhibits almost none.

\Cref{fig:pinch} compares the $\cramped{T=0}$ spin structure factor with the self-consistent Gaussian approximation of the ideal ($\cramped{g=0}$) liquid~\cite{henleyPowerlawSpinCorrelations2005,bentonSpinliquidPinchlineSingularities2016,chung2formU1Spin2025}.
On the pyrochlore lattice ($\cramped{\xig\approx22\,\dnn}$) the pinch points stay sharp at the resolution of $L=128$; on the octochlore lattice ($\cramped{\xig\approx2.6\,\dnn}$) they are visibly smeared, and on the cubichlore lattice ($\cramped{\xig\approx0.9\,\dnn}$) none survives.
Reference~\cite{chung2formU1Spin2025}, using $L=32$, attributed the reduced pinch-point intensity on the pyrochlore lattice to topological sector freezing.
Such sectors do exist, and they affect the intensity at the pinch points themselves~\cite{SM}.
At $L=32$, however, the nearest $\bm q$ points lie outside the smeared region; the four times finer $\bm q$ grid at $L=128$ resolves the smearing around the pinch points, which grows systematically with $|g|$ across the three lattices and so comes from the locking.
On the pyrochlore lattice, the weak height locking allows nearly sharp pinch points to coexist with the magnetic Bragg peaks of PSS order.
Conversely, BSS order has no magnetic Bragg peaks: its signatures are the smeared pinch points, with the lost spectral weight reappearing as broad maxima at the zone center.
This order can be resolved as Bragg peaks in a four-spin correlator of the cluster moments [\cref{app:fourspin}].

\paragraph{Finite-temperature transitions} 
All three lattices undergo a single continuous transition, at $\cramped{T_c/|J_1|=1.6290(6)}$ (octochlore), $1.2271(5)$ (cubichlore), and $1.1565(1)$ (pyrochlore~\cite{chung2formU1Spin2025}), at which the gauge charge strings also proliferate. 
Remarkably, despite sharing their ground-state manifolds with the 3-state Potts antiferromagnets, all three SVMs belong to 3D Ising rather than the 3D XY universality class of the Potts models, confirmed by finite-size scaling with the Ising exponent $\cramped{\nu=0.630}$ [\cref{app:critical}].
The difference follows from the height mapping: the sixfold anisotropy of the Potts order parameter, irrelevant at the XY fixed point~\cite{oshikawaOrderedPhaseScaling2000}, becomes the twofold $\cos(2\varphi)$ in the SVM, as the cyclic color shift becomes a height redundancy.
This twofold anisotropy is strongly relevant at the XY fixed point~\cite{hasenbuschAnisotropicPerturbationsThreedimensional2011}, driving the transition into the Ising class.
For weak locking, however, an extended crossover can mimic XY criticality before the anisotropy becomes dominant.
On the pyrochlore, this is reflected in the non-monotonic drift of the specific heat peak [\cref{fig:T0}(f)] and the Binder cumulants of $\PP$ and $\BB$ [\cref{app:critical}], explaining why the asymptotic Ising behavior can only be resolved for very large system sizes.

\paragraph{Discussion}
Our results reveal a failure of emergence in 3D SVMs: the microscopic discreteness of the spins generates a relevant locking potential which destabilizes the 2-form Coulomb CSL, rendering the gauge fluctuations massive at long distances. 
Nevertheless, characteristic spin-liquid correlations and fractionalized string excitations persist at intermediate scales. 
The ground-state entropy remains extensive despite the order, as locking leaves substantial microscopic freedom.
The pyrochlore provides a particularly striking example: its exceptionally weak locking explains the apparent Coulomb behavior and unresolved criticality reported in Ref.~\cite{chung2formU1Spin2025}. 
The cubichlore and octochlore examples studied here further demonstrate that the locking instability need not appear in conventional magnetic probes, concealing the underlying order.

This mechanism parallels the monopole instability of compact U(1) gauge theories in $\cramped{2\!+\!1}$ dimensions~\cite{polyakovQuarkConfinementTopology1977a,gopfertProofConfinementStatic1982,banksPhaseTransitionsAbelian1977}.
There, the algebraic spin liquid can be stabilized either by adding gapless matter~\cite{hermeleStability1Spin2004,songUnifyingDescriptionCompeting2019}, or at a deconfined quantum critical point (DQCP)~\cite{senthilDeconfinedQuantumCritical2004}. 
For 2-form spin liquids, a potential route to stability is to forbid the $\cos(2\varphi)$ anisotropy via lattice symmetries, leaving $\cos(4\varphi)$ which is irrelevant at the 3D XY fixed point~\cite{shaoMonteCarloRenormalization2020,banerjeeConformalDimensionsLarge2018,hasenbuschAnisotropicPerturbationsThreedimensional2011}. 
The resulting finite-temperature transition could host an emergent 2-form Coulomb theory with algebraic correlations, analogous to a DQCP.

More generally, other candidate spin liquids, especially higher-rank tensor ones~\cite{sternFractonSpinLiquid2026a,niggemannClassicalFractonSpin2026,plackeIsingFractonSpin2024,bentonSpinliquidPinchlineSingularities2016,bentonTopologicalRouteNew2021}, may harbor similar instabilities that could remain hidden if the locking is weak or the order leaves little signature in the spin structure factor. 
While flat bands provide a Gaussian framework for identifying classical spin liquids~\cite{yanClassificationClassicalSpin2024,yanClassificationClassicalSpin2024a,davierCombinedApproachAnalyze2023}, they do not capture the full constraints of the microscopic spins, which can preclude spin liquidity even when the ground-state entropy is extensive and local constraints are satisfied exactly.
Our results provide a complementary perspective on fluctuation-induced ordering in frustrated spin models with flat bands~\cite{YanRank2,franciniHigherrankSpinLiquids2025,chalkerHiddenOrderFrustrated1992,henleyOrderingDueDisorder1989b,villainOrderEffectDisorder1980a,bergmanOrderbydisorderSpiralSpinliquid2007}, exposing a concrete mechanism by which these constraints destabilize the emergent gauge theory.

{\paragraph{Note added} During the final preparation of this manuscript, we became aware of Ref.~\cite{rosalesHiddenEntropicOrder2026}, which independently reports similar results for the octochlore lattice.}

\paragraph{Acknowledgments}
The work at the University of Waterloo was funded by the NSERC of Canada as well as being supported by a University of  Waterloo Research Chair (M.J.P.G.).
M.D.B. acknowledges CGS-M and CGS-D scholarships from NSERC.  
M.U.F. acknowledges a USRA scholarship from NSERC.
K.T.K.C. acknowledges the generous support of the Carl and Lillian Illig Post-Doctoral Fellowship in Nanotechnology,
administered by the Smalley-Curl Institute.
This research was enabled in a large part by computing resources provided by the Digital Research Alliance of Canada. 
The authors acknowledge the use of Claude (Opus 5.5) and ChatGPT (GPT-6) for performance optimization of code originally written and validated by the authors, as well as for literature searches and refinement of clarity and readability of the manuscript. 
The scientific contributions of this work are the authors’ own, and the authors take full responsibility for its content.

\nocite{hasenbuschFiniteSizeScaling2010,campostriniCriticalBehaviorThreedimensional2001,binderFiniteSizeScaling1981,ferrenbergPushingLimitsMonte2018,gliozziCriticalExponents3d2014,hasenbuschEliminatingLeadingSubleading2025,chesterCarvingOutOPE2020}
\bibliography{LowTSVM,SMnote}

\renewcommand{\theequation}{E\arabic{equation}}
\renewcommand{\thefigure}{E\arabic{figure}}
\renewcommand{\thetable}{E\Roman{table}}
\setcounter{equation}{0}
\setcounter{figure}{0}
\setcounter{table}{0}

\clearpage

\onecolumngrid
\section*{End Matter}
\twocolumngrid
\appendix
\renewcommand{\theequation}{\thesection\arabic{equation}} 
\setcounter{secnumdepth}{1} 
\newcommand{\emapp}[2]{\par\refstepcounter{section}\label[appendix]{#2}\textbf{Appendix~\thesection: #1} --- \ignorespaces} 

\emapp{Potts mapping}{app:potts}
Since neighboring heights differ by $\pm1$, the colors $\cramped{c_v=h_v\bmod3}$ of neighboring vertices always differ, so every SVM ground state defines a proper 3-coloring of the parent lattice.
These colorings are ground states of the $\cramped{q=3}$ Potts antiferromagnet, $H_{\textrm P}=J_{\textrm P}\sum_{\langle ab\rangle}\delta_{c_ac_b}$ with $J_{\textrm P}>0$.
Conversely, on the simple-cubic and bcc lattices every proper 3-coloring $c_v$ admits a consistent lift to integer heights $h_v$, uniquely determined by fixing the height at a reference vertex, up to periodic boundary conditions, because the gradients of $c_v$ and $h_v$ coincide ($\pm 1$ mod $3$). 
The global height-shift redundancy $\cramped{h_v \sim h_v + 2}$ then identifies three colorings related by cyclic permutations, establishing a 3-to-1 correspondence between Potts and SVM ground states.
On the diamond lattice, hexagonal loops introduce an obstruction: a hexagon colored $0,1,2,0,1,2$, which we call a chiral hexagon, would require a non-zero spin vorticity $\cramped{\omega_p=\pm6}$. Therefore, the Potts ground states without chiral hexagons, a proper subset of the full Potts ground-state manifold, correspond 3-to-1 to the pyrochlore SVM ground states~\cite{SM}.
These correspondences are exact locally. 
Globally, non-trivial winding sectors  need not map to consistent colorings. See~\cite{SM} for further details.

This correspondence relates the residual entropies of the SVM and Potts models. 
Letting $s_v$ denote the Potts ground-state entropy per parent vertex, the SVM entropy per spin is $\cramped{S_0=2s_v/z}$ (with $z/2$ spins per vertex) in the thermodynamic limit.
This relation is exact on the simple-cubic and bcc lattices, while on the diamond lattice it requires restricting the Potts ground-state ensemble to colorings without chiral hexagons.
We compute $s_v$ for this restricted ensemble directly, by giving each chiral hexagon a fugacity $y$ and integrating from $y=1$ (all colorings) to $y=0$ (no chiral hexagons)~\cite{SM}.
\Cref{tab:entropy} compares these entropies with direct thermodynamic integration of the SVM, finding agreement to within $3\times10^{-5}$ per spin on all three lattices.
The unrestricted diamond ensemble would instead overestimate the pyrochlore entropy by $3\times10^{-4}$ per spin.
Our pyrochlore estimate refines the earlier value from Ref.~\cite{chung2formU1Spin2025}.
The entropy also shows how much freedom the PSS and BSS orders leave: perfect BSS states alone, with one sublattice fixed to a single color and the other choosing freely between the remaining two colors, give $\cramped{s_v=\ln(2)/2}$, and the excess $\cramped{\delta s_v = s_v - \ln(2)/2}$ shrinks as the coordination grows (\cref{tab:entropy}).

\begin{table}[b]
  \caption{Residual entropy per spin $S_0$ of the SVMs compared with the Potts prediction $2s_v/z$; SVM values from Monte Carlo. $^*$Colorings without chiral hexagons. Details are given in the SM~\cite{SM}.}
  \label{tab:entropy}
  \begin{ruledtabular}
  \setlength{\tabcolsep}{1pt}
  \begin{tabular}{lcccc}
    SVM (parent) & Potts $s_v$ & $\delta s_v$ & $2s_v/z$ & SVM $S_0$ \\
    \hline
    pyro (diamond) & 0.41078(2)$^*$ & 0.0642 & 0.20539(1)  & 0.205368(8) \\
    octo (sc)      & 0.36699(3)  & 0.0204 & 0.12233(1)  & 0.122343(8) \\
    cubi (bcc)     & 0.351770(7) & 0.0052 & 0.087943(2) & 0.087948(18) \\
  \end{tabular}
  \end{ruledtabular}
\end{table}

\begin{figure}[b]
  \setlength{\unitlength}{1pt}
  \begin{picture}(246,202)
    \put(0,0){\includegraphics{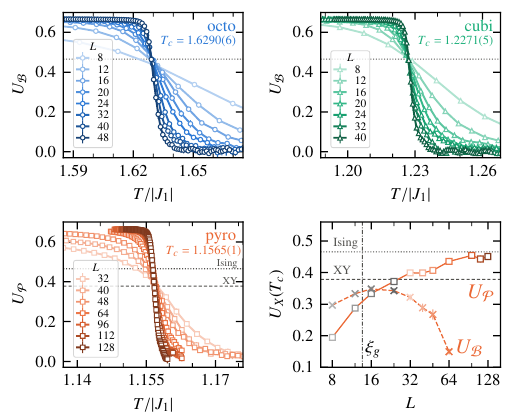}}
    \put(0,194){(a)}\put(124,194){(b)}\put(0,93){(c)}\put(124,93){(d)}
  \end{picture}
   \vspace{-5mm}
  \caption{Finite-temperature transitions. (a,~b) Binder cumulant $U_\BB$ [\cref{eq:binder}] on the octochlore ($L=8$--$48$) and cubichlore ($L=8$--$40$) lattices. Dotted line: the 3D Ising value $\cramped{U^*=0.4655}$~\cite{hasenbuschFiniteSizeScaling2010}; dashed: the 3D XY value $\cramped{U^*\approx0.3785}$ for one component of an O(2) vector~\cite{campostriniCriticalBehaviorThreedimensional2001}; both values are for a periodic cubic box. (c) $U_\PP$ on the pyrochlore lattice ($L=32$--$128$); lines as in (a). (d) $U_\PP$ (squares) and $U_\BB$ ($\times$) on the pyrochlore lattice at $T_c$ against $L$ ($L=8$--$24$ in gray), interpolated linearly between the two simulated temperatures bracketing $T_c$; dash-dotted line: the $T=0$ locking length $L=\xig$.}
  \label{fig:finiteT}
\end{figure}

\newpage

\emapp{Critical behavior}{app:critical}
We locate the transitions with the Binder cumulant of the order parameter~\cite{binderFiniteSizeScaling1981},
\begin{equation}
  U_X = 1-{\langle X^4\rangle}/{3\langle X^2\rangle^2},
  \label{eq:binder}
\end{equation}
with $\cramped{X \equiv \BB}$ on the octochlore and cubichlore lattices and $\cramped{X \equiv \PP}$ on the pyrochlore lattice.
The system size needed to reach the asymptotic critical behavior differs between the lattices: for the octochlore and cubichlore lattices ($\cramped{\xig\lesssim2}$ cells), $U_\BB$ is already in the scaling regime for $\cramped{L\gtrsim16}$ [\cref{fig:finiteT}(a,~b)], whereas the pyrochlore lattice requires $\cramped{L\gtrsim 64}$ [\cref{fig:finiteT}(c)].

\paragraph{XY-to-Ising crossover}
The $\cramped{q=3}$ Potts antiferromagnet on the simple-cubic and bcc lattices has a 3D XY transition~\cite{wangThreestateAntiferromagneticPotts1990,gottlobThreestateAntiferromagneticPotts1994,kurbanovaCriticalPropertiesAntiferromagnetic2025}: its order parameter is an angle with a sixfold anisotropy, which is irrelevant at the XY fixed point, so it pins the angle only once the system has entered the ordered phase~\cite{oshikawaOrderedPhaseScaling2000}.
A cyclic color shift gives the same spin configuration, so the natural SVM angle $\varphi$ is three times the Potts angle and the same anisotropy becomes $\cos(2\varphi)$.
This twofold anisotropy is relevant at the XY fixed point~\cite{hasenbuschAnisotropicPerturbationsThreedimensional2011}, selecting a preferred axis and driving the transition into the Ising universality class. 
The system size required to observe the asymptotic behavior depends on $|g|$: upon cooling, the system first shows XY-like critical behavior and crosses over to Ising when the correlation length exceeds $\cramped{L_\times \sim |u_2|^{-1/Y_2}}$, where $u_2$ is the anisotropy scaling field at $T_c$ (it grows with $|g|$) and $\cramped{Y_2\approx1.76}$ is its RG eigenvalue~\cite{hasenbuschAnisotropicPerturbationsThreedimensional2011}.
The $T=0$ locking length $\xig$ gives a qualitative guide to $L_\times$; the finite-temperature data locate the crossover.
The octochlore and cubichlore lattices are past the crossover at all simulated sizes: their $T=0$ histograms show BSS peaks already at $L=16$ [\cref{fig:T0}(b)], and their $U_\BB$ curves cross cleanly near the Ising value [\cref{fig:finiteT}(a,~b)].
On the pyrochlore lattice the crossover falls within our range of sizes, \emph{even at $T=0$}: the histograms change from a uniform ring at $L=16$ to two PSS peaks for $\cramped{L\gtrsim64\approx5\xig}$ [\cref{fig:T0}(b)], and the $U_\PP$ curves cross close to the Ising value only for $\cramped{L\geq64}$ [\cref{fig:finiteT}(c)].
To understand this behavior, and guided by the $\cramped{T=0}$ histograms of \cref{fig:T0}(b), we track the temperature dependence of both $\BB$ and $\PP$, the two orthogonal components of the O(2) order parameter $e^{i\varphi}$, rather than only the one that eventually orders.
For small system sizes ($L\lesssim L_\times$) the system fluctuates between the two components and their respective Binder cumulants coincide, exhibiting XY-like behavior.
For large system sizes ($L\gtrsim L_\times$) the angle locks: $\PP$ becomes the genuine critical Ising order parameter while $\BB$ becomes a non-critical massive transverse mode.
Accordingly, $U_\BB$ and $U_\PP$ at $T_c$ meet near the XY value at $L\approx18$ ($\approx1.3\,\xig$), and then separate, with $U_\PP$ rising toward the Ising value and $U_\BB$ falling [\cref{fig:finiteT}(d)].
This crossover causes the specific heat peak to drift non-monotonically with $L$ [\cref{fig:T0}(f)]~\cite{SM}.

\begin{table}[t!]
  \caption{Critical temperatures, Binder crossings and exponent ratios of the SVMs, with the 3D Ising~\cite{hasenbuschFiniteSizeScaling2010,ferrenbergPushingLimitsMonte2018} and 3D XY~\cite{campostriniCriticalBehaviorThreedimensional2001} values. Fitting and error details are given in the SM~\cite{SM}.}
  \label{tab:exponents}
  \begin{ruledtabular}
  \setlength{\tabcolsep}{1pt}
  \footnotesize
  \begin{tabular}{lcccccc}
    & $T_c/|J_1|$ & $U^*$ & $\beta/\nu$ & $\gamma/\nu$ & $1/\nu$ & $(2\beta+\gamma)/\nu$ \\
    \hline
    pyro & 1.1565(1) & 0.44(1)   & 0.56(5)   & 1.92(5)   & 1.54(13) & 3.04(13) \\
    octo & 1.6290(6) & 0.481(11) & 0.513(21) & 1.987(20) & 1.60(3) & 3.01(4) \\
    cubi & 1.2271(5) & 0.446(12) & 0.531(24) & 1.953(36) & 1.67(3) & 3.02(1) \\
    \hline
    3D Ising & & 0.4655 & 0.518 & 1.964 & 1.587 & 3 \\
    3D XY    & & 0.3785 & 0.519 & 1.962 & 1.488 & 3 \\
  \end{tabular}
  \end{ruledtabular}
   \vspace{-4mm}
\end{table}

\paragraph{Universality class}
All three transitions belong to the 3D Ising class. The Ising and XY classes are challenging to separate because $\beta/\nu$ and $\gamma/\nu$ agree within $0.2\%$. 
They mainly differ in $\nu$ and in the sign of $\cramped{\alpha=2-d\nu}$ ($d=3$). 
We use these two differences to identify the universality~\cite{gliozziCriticalExponents3d2014,hasenbuschEliminatingLeadingSubleading2025,chesterCarvingOutOPE2020}.
First, the specific heat maxima grow with $L$ as $L^{\alpha/\nu}$ with $\cramped{\alpha/\nu\approx0.2}$--$0.3$, consistent with a positive $\alpha$; in the 3D XY class $\alpha$ is negative and the peak approaches a finite cusp.
Second, two estimates of $\nu$ on the octochlore and cubichlore lattices, from the growth of the maximum slope of $U_\BB$ and from a scaling collapse of $U_\BB$, both give $1/\nu\approx1.6$, close to the Ising value $1.587$ and away from the XY value $1.488$ [\cref{tab:exponents}].
The remaining exponents are consistent with this, and the Binder crossings provide a further consistency check; since $U^*$ depends on the shape of the simulation box, we do not use it to identify the class~\cite{SM}.
For the pyrochlore lattice, only $\cramped{L=64}$--$128$ are in the scaling regime, so its exponents carry larger errors, and the assertion of Ising critical behavior rests mainly on the system size evolution of $U_\PP$ described above.

A single lattice-dependent coupling $g$ thus organizes the transitions for all three lattices.
Its sign selects which $\mathbb{Z}_2$ symmetry breaks: the $A\leftrightarrow B$ sublattice symmetry for $\BB$ (BSS) or time reversal for $\PP$ (PSS).
Its magnitude, which varies over three orders of magnitude, sets how large a system must be before the Ising critical behavior appears.
For large enough systems, all three SVMs undergo a 3D Ising transition.

\begin{figure}[b]
  \setlength{\unitlength}{1pt}
  \begin{picture}(246,197)
    \put(0,0){\includegraphics{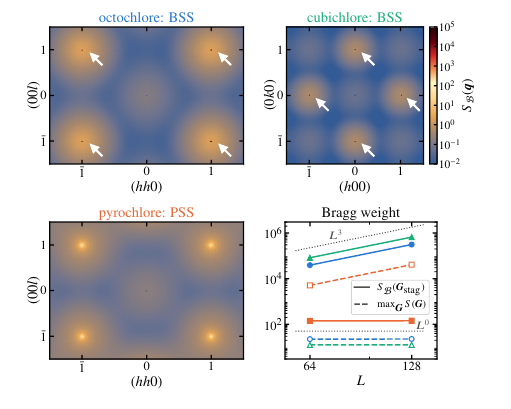}}
    \put(0,187){(a)}\put(118,187){(b)}\put(0,93){(c)}\put(118,93){(d)}
  \end{picture}
   \vspace{-8mm}   
 \caption{Broken-sublattice (BSS) order in the four-point correlator at $\cramped{T=0}$. (a--c) Cluster structure factor $S_\BB(\bm q)$ at $\cramped{L=64}$ for all lattices. Sharp Bragg peaks (black dots) appear at the staggering wave vectors---$\cramped{(\pm1\pm\!1\pm\!1)}$ in (a), $\cramped{(\pm100)}$ and $\cramped{(0\pm\!10)}$ in (b)---indicated by white arrows. The pyrochlore lattice (c) has none. (d) Raw Bragg weights against $L$: $S_\BB(\bm G_{\textrm{stag}})$ (filled, solid) and the largest spin Bragg weight $\max_{\bm G}S(\bm G)$ (open, dashed); dotted guides: $L^3$ and $L^0$.}
  \label{fig:fourspin}
\end{figure}

\emapp{Four-point correlator of BSS order}{app:fourspin}
In an ideal BSS state [\cref{fig:BSS_PSS}] one sublattice has random AIAO moments $m_v = \pm 1$, giving short-range spin-spin correlations within a cluster but not beyond it.
The spin structure factor $S(\bm q)$ therefore shows only diffuse intensity and broad zone-center peaks. 
The order appears instead in the correlations of the squared cluster moments $\cramped{Q_v=m_v^2}$, and can be diagnosed with a four-spin correlation structure factor,
\begin{equation}
  S_\BB(\bm q)=\frac{1}{N_v}\Big\langle\Big|\sum_v
  [Q_v - \bar Q]
  e^{-i\bm q\cdot\bm x_v}\Big|^2\Big\rangle,
  \label{eq:SB}
\end{equation}
where the sum runs over the $N_v=2L^3$ parent vertices at positions $\bm x_v$, and $\bar Q=N_v^{-1}\sum_v Q_v$ is the spatial mean in each configuration.
BSS order makes $Q_v$ larger on one sublattice than the other, so $S_\BB$ develops Bragg peaks at the wave vectors $\bm G_{\textrm{stag}}$ that distinguish the A and B sublattices [\cref{fig:fourspin}(a,b)],
scaling as $L^3$ [\cref{fig:fourspin}(d)].
On the pyrochlore lattice the roles are reversed: the PSS (AIAO) order gives spin Bragg peaks in $S(\bm q)$ but not $S_{\BB}(\bm q)$~[\cref{fig:fourspin}(c)].
Since PSS treats both sublattices alike, the weight of $S_\BB(\bm q)$ at $\bm G_{\textrm{stag}}$ does not grow with $L$ [\cref{fig:fourspin}(d)].
Each of the two orders is therefore observed in only one of the two correlators: PSS in $S(\bm q)$ and BSS in $S_\BB(\bm q)$.

\onecolumngrid
\clearpage
\twocolumngrid

\renewcommand{\theequation}{S\arabic{equation}}
\renewcommand{\thefigure}{S\arabic{figure}}
\renewcommand{\thetable}{S\Roman{table}}
\setcounter{equation}{0}
\setcounter{figure}{0}
\setcounter{table}{0}

\onecolumngrid
\begin{center}
    {\large\bfseries
    Supplemental Material for
    ``\papertitle''\par}

    \vspace{1em}

    {\normalsize
    \authorMichael$^{1,*}$,
    \authorMohid$^{1,*}$,
    \authorKai$^{2}$, and
    \authorMichel$^{1,3}$\par}

    \vspace{0.5em}

    {\small\itshape
    $^{1}$\affiliationWaterloo\\
    $^{2}$\affiliationRice\\
    $^{3}$\affiliationNano\par}

    \vspace{0.5em}

    {\small
    $^{*}$These authors contributed equally to this work.\par}

    \vspace{1em}
\end{center}
\twocolumngrid

\newcounter{smsec}\renewcommand{\thesmsec}{S\arabic{smsec}}
\newcommand{\smsection}[2]{\refstepcounter{smsec}\label{#2}\section*{\thesmsec.\ #1}}

\smsection{Lattice geometry and Fourier-space coupling matrices}{sm:Jq}
This section describes the conventions used throughout the manuscript: where the spins sit and how their Ising axes point, the spin-spin couplings obtained by expanding \cref{eq:H}, and the Fourier-space coupling matrices, whose two flat bands underlie the extensive ground-state degeneracy.

\paragraph{Lattice geometry}
Each lattice is a Bravais lattice with primitive vectors $\bm a_{1,2,3}$ and a basis of $z$ spins, one cluster per sublattice per cell.
Measured from the center of an A cluster, spin $\mu$ sits at $\bm c_\mu$, the midpoint of the parent bond to the B cluster at $2\bm c_\mu$; its Ising axis is $\hat{\bm e}_\mu=\bm c_\mu/|\bm c_\mu|$, so $\sigma_\mu=+1$ points from A to B.
In units of the conventional cubic lattice constant $a$:

\textit{Pyrochlore} (fcc Bravais lattice; diamond parent, $\dnn=\sqrt3\,a/4$):
\begin{align}
  \bm a_1&=\tfrac a2(\hat{\bm x}+\hat{\bm y}),\ \ \bm a_2=\tfrac a2(\hat{\bm x}+\hat{\bm z}),\ \ \bm a_3=\tfrac a2(\hat{\bm y}+\hat{\bm z}),\nonumber\\
  \bm c_1&=\tfrac a8(-\hat{\bm x}-\hat{\bm y}-\hat{\bm z}),\ \ \bm c_2=\tfrac a8(\hat{\bm x}+\hat{\bm y}-\hat{\bm z}),\nonumber\\
  \bm c_3&=\tfrac a8(\hat{\bm x}-\hat{\bm y}+\hat{\bm z}),\ \ \bm c_4=\tfrac a8(-\hat{\bm x}+\hat{\bm y}+\hat{\bm z}).
\end{align}

\textit{Octochlore} (fcc Bravais lattice; simple-cubic parent, $\dnn=a/2$):
\begin{align}
  \bm a_1&=\tfrac a2(\hat{\bm x}+\hat{\bm y}),\ \ \bm a_2=\tfrac a2(\hat{\bm x}+\hat{\bm z}),\ \ \bm a_3=\tfrac a2(\hat{\bm y}+\hat{\bm z}),\nonumber\\
  \bm c_1&=\tfrac a4\hat{\bm x},\ \ \bm c_2=\tfrac a4\hat{\bm y},\ \ \bm c_3=\tfrac a4\hat{\bm z},\ \ \bm c_{\mu+3}=-\bm c_\mu\ \ (\mu\bmod6).
\end{align}

\textit{Cubichlore} (simple-cubic Bravais lattice; bcc parent, $\dnn=\sqrt3\,a/2$):
\begin{align}
  \bm a_1&=a\hat{\bm x},\ \ \bm a_2=a\hat{\bm y},\ \ \bm a_3=a\hat{\bm z},\nonumber\\
  \bm c_\mu&=\tfrac a4\big(s_\mu^x\hat{\bm x}+s_\mu^y\hat{\bm y}+s_\mu^z\hat{\bm z}\big),
\end{align}
with $(s_\mu^x,s_\mu^y,s_\mu^z)=$ $({-}{-}{-})$, $({+}{-}{-})$, $({-}{+}{-})$, $({+}{+}{-})$, $({-}{-}{+})$, $({+}{-}{+})$, $({-}{+}{+})$, $({+}{+}{+})$ for $\mu=1,\dots,8$.
This numbering of the spins is used for all matrices, including those below.
A system of linear size $L$ contains the $L^3$ cells $\bm R=i\bm a_1+j\bm a_2+k\bm a_3$, $i,j,k=0,\dots,L-1$, with periodic boundaries, i.e.\ $N=zL^3$ spins.
For the pyrochlore and octochlore lattices this periodic box is rhombohedral, for the cubichlore lattice it is cubic.
Wavevectors are quoted as $\bm q=(2\pi/a)(h\hat{\bm x}+k\hat{\bm y}+l\hat{\bm z})$, i.e.\ $(h,k,l)$ are Cartesian components in units of $2\pi/a$, not components along the primitive reciprocal lattice vectors $\bm b_i$, defined by $\cramped{\bm a_i\cdot\bm b_j=2\pi\delta_{ij}}$.
The periodic box of $L^3$ cells allows $\cramped{\bm q=\sum_i n_i\bm b_i/L}$ with integer $n_i$.

\begin{figure}[b]
  \setlength{\unitlength}{1pt}
  \begin{picture}(246,90)
    \put(0,0){\includegraphics{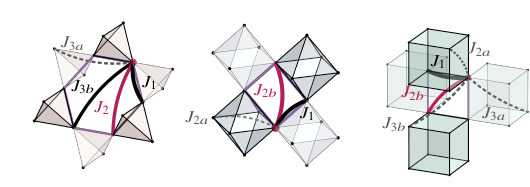}}
    \put(0,82){(a)}\put(83,82){(b)}\put(166,82){(c)}
  \end{picture}
  \caption{Ising couplings $J_n$ of \cref{tab:couplings} on the plaquette clusters: (a) pyrochlore, (b) octochlore, (c) cubichlore. Curves join one reference spin (red) to a representative neighbor in each shell. Colors indicate signs of couplings in the respective SVM Hamiltonian; black: $J_n<0$ (ferromagnetic), magenta: $J_n>0$ (antiferromagnetic), dashed gray: $J_n=0$.}
  \label{fig:couplings}
\end{figure}

\paragraph{Ising couplings}\label{app:lattices}
Expanding $\omega_p^2$ in \cref{eq:H} couples every pair of spins that share a plaquette: hexagons of the diamond parent on the pyrochlore lattice, squares of the simple-cubic parent on the octochlore lattice, and squares of the bcc parent on the cubichlore lattice.
We label the couplings by shell: $J_n$ couples $n$th-nearest neighbors, and neighbors at the same distance that are not related by a lattice symmetry are distinguished by a letter ($J_{2a}$, $J_{2b}$, \dots).
The resulting Ising couplings, with $J_1=-1$, are listed in \cref{tab:couplings} and drawn in \cref{fig:couplings}.

\begin{table}[t]
  \caption{Ising couplings of the three SVMs, from \cref{eq:H} with $J_1=-1$ (units of $|J_1|$). $r$: pair separation in units of the conventional cubic lattice constant $a$; $n$: number of such neighbors of a spin. Shells at equal $r$ that are not related by symmetry carry a letter.}
  \label{tab:couplings}
  \begin{ruledtabular}
  \begin{tabular}{llccr}
    lattice (plaquette) & shell & $r/a$ & $n$ & $J$ \\
    \hline
    pyrochlore   & $J_1$ & $\sqrt2/4$ & 6 & $-1$ \\
    (hexagon)    & $J_2$ & $\sqrt6/4$ & 12 & $+\tfrac12$ \\
                 & $J_{3a}$ (through a shared spin) & $\sqrt2/2$ & 6 & $0$ \\
                 & $J_{3b}$ (across a hexagon) & $\sqrt2/2$ & 6 & $-\tfrac12$ \\
    \hline
    octochlore   & $J_1$ & $\sqrt2/4$ & 8 & $-1$ \\
    (square)     & $J_{2a}$ (same octahedron) & $1/2$ & 2 & $0$ \\
                 & $J_{2b}$ (across a square) & $1/2$ & 4 & $+1$ \\
                 & $J_3$ & $\sqrt6/4$ & 16 & $0$ \\
    \hline
    cubichlore   & $J_1$ & $1/2$ & 6 & $-1$ \\
    (square)     & $J_{2a}$ (same cube) & $\sqrt2/2$ & 6 & $0$ \\
                 & $J_{2b}$ (across a square) & $\sqrt2/2$ & 6 & $+\tfrac12$ \\
                 & $J_{3a}$ (same cube) & $\sqrt3/2$ & 2 & $0$ \\
                 & $J_{3b}$ & $\sqrt3/2$ & 6 & $0$ \\
  \end{tabular}
  \end{ruledtabular}
\end{table}

\begin{figure}[t]
  \setlength{\unitlength}{1pt}
  \begin{picture}(246,163)
    \put(0,0){\includegraphics{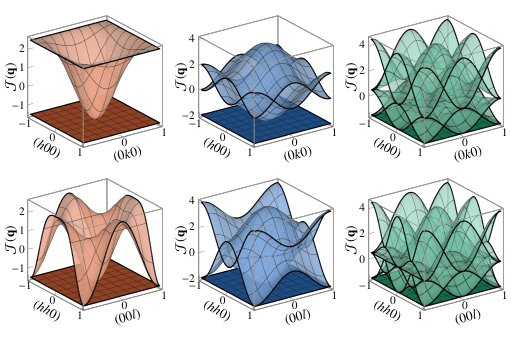}}
    \put(0,157){(a)}\put(82,157){(b)}\put(164,157){(c)}
  \end{picture}
  \caption{Eigenvalues of the coupling matrix $\mathcal J(\bm q)$ [\cref{eq:Jq}], in units of $|J_1|$, for the (a)~pyrochlore, (b)~octochlore and (c)~cubichlore SVMs, on the $(hk0)$ plane (top) and the $(hhl)$ plane (bottom), with $\bm q=2\pi(h,k,l)/a$. Solid sheets: the two degenerate flat bands at the minimum ($-\tfrac32$, $-2$ and $-\tfrac32$); translucent sheets: the $z-2$ dispersive bands.}
  \label{fig:Jqbands}
\end{figure}

\paragraph{Fourier transform and flat bands}
We use the Fourier convention of Ref.~\cite{sternFractonSpinLiquid2026a}:
\begin{equation}
  \sigma_{\bm R\mu}=\frac{1}{\sqrt{L^3}}\sum_{\bm q}e^{i\bm q\cdot(\bm R+\bm c_\mu)}\,\sigma_\mu(\bm q).
\end{equation}
Up to a constant, \cref{eq:H} is $H=\sum_n J_n\sum_{\langle ij\rangle_n}\sigma_i\sigma_j$ with the couplings of \cref{tab:couplings}, so that
\begin{gather}
  H=\sum_{\bm q}\sum_{\mu\nu}\mathcal J_{\mu\nu}(\bm q)\,\sigma_\mu(\bm q)\,\sigma_\nu(-\bm q),\nonumber\\
  \mathcal J(\bm q)=\frac12\sum_n J_n \mathcal A^{(n)}(\bm q),\nonumber\\
  \mathcal A^{(n)}_{\mu\nu}(\bm q)=\sum_{\bm R}A^{(n)}_{(\bm 0\mu),(\bm R\nu)}\,e^{i\bm q\cdot(\bm R+\bm c_\nu-\bm c_\mu)},
  \label{eq:Jq}
\end{gather}
where $\cramped{A^{(n)}_{(\bm 0\mu),(\bm R\nu)}\in\{0,1\}}$ is the adjacency matrix of shell $n$.
The $\mathcal A^{(n)}(\bm q)$ are real and symmetric; they are listed below.
On every lattice the lowest eigenvalue of $\mathcal J(\bm q)$ [$-\tfrac32$, $-2$ and $-\tfrac32$ in units of $|J_1|$ for the pyrochlore, octochlore and cubichlore lattices] is twofold degenerate at generic $\bm q$: two flat bands lie below $z-2$ dispersive ones [\cref{fig:Jqbands}].
The flat bands are the configurations with $\omega_p=0$ on every plaquette.
Such configurations are lattice gradients of a height field, $\sigma_{\langle ab\rangle}=h_b-h_a$ [\cref{eq:height}].
Because the heights live on the two sublattices A and B, the flat bands at each $\bm q$ are spanned by two vectors, $\bm g_A$ and $\bm g_B$, with components $(\bm g_A)_\mu=-e^{-i\bm q\cdot\bm c_\mu}$ and $(\bm g_B)_\mu=e^{i\bm q\cdot\bm c_\mu}$ for $\mu=1,\dots,z$.
They are the spin patterns produced by raising the height of a single A or B vertex, written in Fourier space.
The extensive ground-state manifold of the Ising model is the set of integer height fields compatible with $\sigma=\pm1$.

Here we list the matrices $\mathcal A^{(n)}(\bm q)$ of \cref{eq:Jq} for the shells with $J_n\neq0$ (\cref{tab:couplings}), with the spins numbered as above.
We write $c_\alpha=\cos(q_\alpha a/2)$, $c_{\alpha\beta}=\cos[(q_\alpha+q_\beta)a/4]$, $c_{\alpha\bar\beta}=\cos[(q_\alpha-q_\beta)a/4]$, $\tilde c_{\alpha\beta}=\cos[(q_\alpha+q_\beta)a/2]$ and $\tilde c_{\alpha\bar\beta}=\cos[(q_\alpha-q_\beta)a/2]$.

\textit{Pyrochlore}, $\mathcal J=\tfrac12\big[-\mathcal A^{(1)}+\tfrac12\mathcal A^{(2)}-\tfrac12\mathcal A^{(3b)}\big]$:
\begin{align}
  \mathcal A^{(1)}&=2\begin{pmatrix}
    0&c_{xy}&c_{xz}&c_{yz}\\
    c_{xy}&0&c_{y\bar z}&c_{x\bar z}\\
    c_{xz}&c_{y\bar z}&0&c_{x\bar y}\\
    c_{yz}&c_{x\bar z}&c_{x\bar y}&0
  \end{pmatrix},\\
  \mathcal A^{(2)}&=4\begin{pmatrix}
    0&c_zc_{x\bar y}&c_yc_{x\bar z}&c_xc_{y\bar z}\\
    c_zc_{x\bar y}&0&c_xc_{yz}&c_yc_{xz}\\
    c_yc_{x\bar z}&c_xc_{yz}&0&c_zc_{xy}\\
    c_xc_{y\bar z}&c_yc_{xz}&c_zc_{xy}&0
  \end{pmatrix},
\end{align}
and $\mathcal A^{(3b)}$ is diagonal, $\mathcal A^{(3b)}_{11}=2(\tilde c_{x\bar y}+\tilde c_{x\bar z}+\tilde c_{y\bar z})$, $\mathcal A^{(3b)}_{22}=2(\tilde c_{x\bar y}+\tilde c_{xz}+\tilde c_{yz})$, $\mathcal A^{(3b)}_{33}=2(\tilde c_{xy}+\tilde c_{x\bar z}+\tilde c_{yz})$, $\mathcal A^{(3b)}_{44}=2(\tilde c_{xy}+\tilde c_{xz}+\tilde c_{y\bar z})$.

\textit{Octochlore}, $\mathcal J=\tfrac12\big[-\mathcal A^{(1)}+\mathcal A^{(2b)}\big]$, in $3\times3$ blocks ($\mu=1,2,3$ and $4,5,6$):
\begin{gather}
  \mathcal A^{(1)}=2\begin{pmatrix}M_-&M_+\\M_+&M_-\end{pmatrix},
  \qquad
  \mathcal A^{(2b)}=2\begin{pmatrix}0&D\\D&0\end{pmatrix},\nonumber\\
  M_\pm=\begin{pmatrix}
    0&c_{x\pm y}&c_{x\pm z}\\
    c_{x\pm y}&0&c_{y\pm z}\\
    c_{x\pm z}&c_{y\pm z}&0
  \end{pmatrix},
  \\
  D= \begin{pmatrix}
      c_y + c_z & 0 & 0 \\ 0 & c_x+c_z & 0  \\ 0 & 0 & c_x + c_y
  \end{pmatrix},
\end{gather}
where $c_{x+y}\equiv c_{xy}$ and $c_{x-y}\equiv c_{x\bar y}$.

\textit{Cubichlore}, $\mathcal J=\tfrac12\big[-\mathcal A^{(1)}+\tfrac12\mathcal A^{(2b)}\big]$, in $4\times4$ blocks ($\mu=1$--$4$, with $s^z_\mu=-1$, and $\mu=5$--$8$, with $s^z_\mu=+1$).
The eight spins sit at the corners of a cube: $\mathcal A^{(1)}$ connects spins whose signs $s_\mu$ differ in one component (a cube edge), and $\mathcal A^{(2b)}$ spins whose signs differ in two:
\begin{gather}
\mathcal A^{(1)}=2\begin{pmatrix}B&c_z\openone\\c_z\openone&B\end{pmatrix},
\qquad
\mathcal A^{(2b)}=2\begin{pmatrix}X&Y\\Y^{T}&X\end{pmatrix},\nonumber\\
B=\begin{pmatrix}
0&c_x&c_y&0\\
c_x&0&0&c_y\\
c_y&0&0&c_x\\
0&c_y&c_x&0
\end{pmatrix},\quad
X=\begin{pmatrix}
0&0&0&\tilde c_{x\bar y}\\
0&0&\tilde c_{xy}&0\\
0&\tilde c_{xy}&0&0\\
\tilde c_{x\bar y}&0&0&0
\end{pmatrix},\nonumber\\
Y=\begin{pmatrix}
0&\tilde c_{x\bar z}&\tilde c_{y\bar z}&0\\
\tilde c_{xz}&0&0&\tilde c_{y\bar z}\\
\tilde c_{yz}&0&0&\tilde c_{x\bar z}\\
0&\tilde c_{yz}&\tilde c_{xz}&0
\end{pmatrix}.
\end{gather}

\begin{table*}[htbp]
    \centering
    \renewcommand{\arraystretch}{1.5}
    \begin{ruledtabular}
    \begin{tabular}{llcccccccl}
    type & typical ensemble & $\bar{h}$ & $\varphi$ & $\langle m_A \rangle$ &  $\langle m_A^2 \rangle$ & $\langle m_B^2 \rangle$ & $\PP$ & $\BB$ & spin configuration\\
    \hline
    PSS & $(h_A=0, h_B=1)$ & $1/2$ & $\pi/2$ & $1$ & $1$ & $1$ & $+1$ & $0$ & AIAO, spins point $A \to B$
    \\
    PSS & $(h_A=2, h_B=1)$ & $3/2$ & $3\pi/2$ & $-1$ & $1$ & $1$ & $-1$ & $0$ & AIAO, spins point $B \to A$
    \\
    BSS & $(h_A=0, h_B\in\{-1,1\})$ & $0$ & $0$ & $0$ & $1/z$ & $1$ & $0$ & $\frac{z-1}{z}>0$ & $B$ sublattice randomly all-in or all-out
    \\
    BSS & $(h_A\in\{0,2\}, h_B=1)$ & $1$ & $\pi$ & $0$ & $1$ & $1/z$ & $0$ & $\frac{1-z}{z}<0$ & $A$ sublattice randomly all-in or all-out
    \end{tabular}
    \end{ruledtabular}
    \caption{
    Representative PSS and BSS height ensembles and their corresponding microscopic order parameters. 
    The height average $\bar h$, angle $\varphi=\pi\bar h$, and local moments illustrate the identification $\BB\propto\cos\varphi$ and $\PP\propto\sin\varphi$. 
    The BSS values assume independently and equally distributed heights on the fluctuating sublattices and do not account for short-range correlations.
    These ensembles can be directly compared to the illustrations in \cref{fig:BSS_PSS} in the main text.
    }
    \label{tab:BSS_PSS_examples}
\end{table*}

\smsection{Correlations and Gaussian Approximation}{sm:scga}

We denote the Ising spin-spin correlation function as $\mathcal{G}_{ij} \equiv \langle \sigma_i \sigma_j \rangle$ where angled brackets denote thermal averages and $i,j$ are lattice sites. 
We define the spin structure factor $S(\bm q)$ to be its Fourier transform (defined as in \cref{eq:Jq}) as
\begin{equation}
    S(\bm q) = \frac{1}{zL^3} \sum_{i,j} \langle \sigma_i \sigma_j\rangle \, e^{-i \bm{q} \cdot (\bm{x}_j - \bm{x}_i)} = \frac{1}{z}\sum_{\mu\nu} \mathcal{G}_{\mu\nu}(\bm q).
\end{equation}
The left halves of the maps in \cref{fig:pinch} show what the spin structure factor would look like in the ideal Coulomb phase, without any locking ($g=0$).
We compute it in the self-consistent Gaussian approximation (SCGA)~\cite{sternFractonSpinLiquid2026a}, which replaces the hard constraint $\sigma_\mu=\pm1$ by the softer $\langle\sigma_\mu^2\rangle=1$.
The spin correlations are then Gaussian, $\mathcal G_{\mu\nu}(\bm q)=[\lambda+2\beta\mathcal J(\bm q)]^{-1}_{\mu\nu}$, with $\lambda$ chosen to satisfy the constraint.
As $T\to0$ only the two flat bands of \cref{sm:Jq} survive, and $\mathcal G(\bm q)$ becomes $(z/2)$ times the projector $P(\bm q)$ onto the flat-band space spanned by $\bm g_A$ and $\bm g_B$, so that
\begin{equation}
S^{T=0}_{\textrm{SCGA}}(\bm q)=\frac12\sum_{\mu\nu}P_{\mu\nu}(\bm q).
\end{equation}
This obeys the same sum rule as the Monte Carlo $S(\bm q)$, so the two halves of each map can be compared directly.
We evaluate the SCGA on the same $\bm q$ grid as the $\cramped{L=128}$ Monte Carlo data.

\smsection{Height Representation, Potts Mapping, and Order Parameters}{sm:heights_potts}

This section provides further details of the microscopic height mapping, the correspondence between SVM ground states and those of the 3-state Potts antiferromagnet, and definitions of the microscopic order parameters and their correspondence with the height theory observables.

\paragraph{Height mapping} 
In a ground state every plaquette has $\omega_p=0$. This local constraint is solved exactly by writing the spins as the gradient of an integer height field on the vertices of the parent lattice, $\sigma_{\langle ab\rangle}=h_b-h_a$ with $a\in A$ and $b\in B$ [\cref{eq:height}]~\cite{chung2formU1Spin2025,kondevFourcoloringModelSquare1995,henleyCoulombPhaseFrustrated2010}.
To construct it, fix $\cramped{h_{v_0}=0}$ on one A vertex and set 
\begin{equation}
    h_v=\sum_{\langle ab\rangle\in\Gamma}\eta_{ab}\,\sigma_{\langle ab\rangle}
\end{equation} 
for a path $\Gamma$ of parent bonds from $v_0$ to $v$, where $\eta_{ab}=+1$ ($-1$) for a bond traversed from A to B (B to A). Every bond changes $h$ by $\pm1$, so $h$ is even on A and odd on B. Only height differences define the spins, so $h$ is defined up to a uniform shift; shifts by an even integer preserve this parity.
For $h$ to be well defined, $h_v$ must not depend on $\Gamma$. Two paths to the same vertex together form a closed loop, so it suffices that the change of $h$ around every closed loop vanishes. A contractible loop is the boundary of a set of plaquettes. Summing $\omega_p$ over these plaquettes, every interior bond is traversed once in each direction and cancels, which leaves exactly the change of $h$ around the loop. In a ground state every $\omega_p$ vanishes, so $h$ is single valued on any simply connected region.

On the periodic $L^3$ box the height need not be single valued. Let $\Gamma_i$ be a path of parent bonds that wraps once around the box along $\bm a_i$, from a vertex $v$ to its periodic image $v+L\bm a_i$. Such a loop bounds no set of plaquettes, so the argument above does not apply, and the winding
\begin{equation}
  W_i=\sum_{\langle ab\rangle\in\Gamma_i}\eta_{ab}\,\sigma_{\langle ab\rangle}
\end{equation}
need not vanish. The height then obeys $h_{v+L\bm a_i}=h_v+W_i$, so $W_i/L$ is the average tilt of $h$ along $\bm a_i$. Since $v$ and its image lie on the same sublattice, $W_i$ is even, and since two such paths differ by a contractible loop, $W_i$ depends only on the configuration, not on the choice of $\Gamma_i$.
The windings cannot be changed by local moves between ground states. A move that changes spins only inside a region that does not wrap around the box leaves every $W_i$ unchanged, because $\Gamma_i$ can be chosen to avoid that region; star moves (below), which shift the height of a single vertex by $\pm2$, are an example. The ground-state manifold therefore splits into topological sectors labeled by $(W_1,W_2,W_3)$~\cite{chung2formU1Spin2025}; since every chain along $\bm a_i$ carries the same $W_i$, these are equivalently sectors of fixed net magnetization, $M_i\propto L^2W_i$, analogous to the flux sectors of spin ice~\cite{henleyCoulombPhaseFrustrated2010,jaubertTopologicalSectorFluctuations2013}. At nonzero temperature, single spin flips leave the ground-state manifold, and the sectors are connected through excited states with $\omega_p\neq0$. All our $T=0$ runs start in, and remain in, the sector $W_i=0$. Finite-temperature runs with single-spin flips are not confined to a sector; we have checked that at low $T$ they relax to $W_i=0$, and their order parameters agree with the $T=0$ values. Fixing the sector fixes the uniform tilt of the heights, which is the net magnetization. This mode contributes to $S(\bm q)$ only at the pinch points $\bm q=\bm G$; at other $\bm q$ a fixed sector changes $S(\bm q)$ only by finite-size corrections of order $1/N$. The smearing in \cref{fig:pinch} instead extends over a width $1/\xig$ around each $\bm G$: at $L=128$, two points $\bm q\neq\bm G$ of the pyrochlore cut lie inside $|\delta q|<1/\xig$. This width is set by $\xig$ from the height fits and grows with $|g|$, whereas a sector effect would scale with $1/L$ and look the same on all three lattices.

\paragraph{Potts mapping} 
As described in the main text and in \cref{app:potts}, the height representation also establishes a correspondence with the ground states of the three-state Potts antiferromagnet on the parent lattice. 
Assigning each vertex a color $c_v=h_v\bmod 3$ gives a proper 3-coloring, since neighboring heights differ by $\pm 1$. 
Conversely, a proper coloring uniquely determines the spins through $\sigma_{\langle ab\rangle}=\pm1$, according to whether $c_b-c_a=\pm1\bmod 3$. 
These spins satisfy the SVM ground-state constraint provided their oriented sum vanishes around every elementary plaquette. 
This condition is automatic for square plaquettes, but on the diamond lattice a hexagon with cyclic coloring $0,1,2,0,1,2$ (or its reverse) produces a nonzero circulation $\omega_p=\pm6$, obstructing the reconstruction of a consistent height field. 
Excluding such chiral hexagons therefore gives a three-to-one correspondence between admissible Potts colorings and SVM ground states, since the three global cyclic permutations of the colors yield the same spins. 
Under periodic boundary conditions, the correspondence is restricted to SVM sectors with $W_i=0\bmod 3$ for every direction (or equivalently $W_i=0\bmod 6$ because bipartiteness forces $W_i$ to be even); other sectors have tilted heights whose colors cannot be made periodic. 
These global restrictions do not affect the bulk thermodynamics.

\paragraph{Order parameter definitions}
The microscopic order parameters $\BB$ and $\PP$ of \cref{eq:BP} correspond to the real and imaginary components of the height order parameter $e^{i\varphi}$, respectively. 
This identification is consistent with their symmetry properties: time reversal sends $\cramped{h\to-h}$ and hence $\cramped{\varphi\to-\varphi}$, under which $\BB$ is even and $\PP$ is odd, while sublattice exchange shifts $\cramped{\varphi\to\varphi+\pi}$ and reverses both order parameters. 
Their relative signs can be established from representative locked height configurations as given in \cref{tab:BSS_PSS_examples}. 
For half-odd-integer locking (PSS), uniform spin polarization gives $\PP\propto\sin\varphi$, with its sign distinguishing the two AIAO states.
For integer locking (BSS), one sublattice has saturated local moments while the other fluctuates, so the sublattice imbalance of squared moments gives $\BB\propto\cos\varphi$, with its sign distinguishing which sublattice is locked.
\Cref{tab:BSS_PSS_examples} illustrates these correspondences through the heights, angles, local moments, and resulting order parameters.

\smsection{Height Field Monte Carlo Analysis}{sm:heights_mc}

This section explains how we extract the stiffness $K$ and the locking strength $g$ from the height fluctuations, and describes the star moves, the local moves between ground states used in our $T=0$ Monte Carlo.

\begin{figure}[b]
  \setlength{\unitlength}{1pt}
  \begin{picture}(246,162)
    \put(0,0){\includegraphics{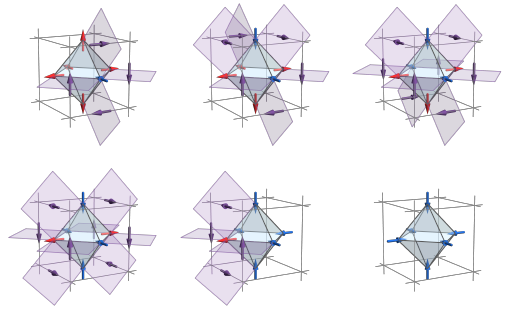}}
    \put(0,154){(a)}\put(83,154){(b)}\put(166,154){(c)}
    \put(0,75){(d)}\put(83,75){(e)}\put(166,75){(f)}
  \end{picture}
  \caption{Star move on one octahedral cluster of the octochlore lattice, adapted from Ref.~\cite{burkeSpinVorticityModels}. (a--e) Reversing one to five of its six spins (blue: reversed, red: not yet reversed) leaves plaquettes with $\omega_p\neq0$ (shaded purple), each marked by its curl vector $\omega_p\hat{\bm n}_p$ (right-hand rule) on an edge of the dual cubic lattice (gray). These curl vectors join into closed strings on the dual lattice, the vortex-string excitations of the SVM. (f) Reversing all six takes the all-out cluster to all-in without violating any plaquette, shifting its height by $\pm2$.}
  \label{fig:stars}
\end{figure}

\paragraph{Star moves}
A single spin flip in a ground state violates the zero-curl constraint on every plaquette containing that spin. Reversing all $z$ spins of a fully polarized (all-in or all-out) cluster instead violates none. In the height language the cluster's vertex is then a local extremum, with all $z$ neighbors at the same height $h_v\mp1$, and the move takes $h_v\to h_v\mp2$ [\cref{fig:stars}]; in Potts language it recolors a vertex whose neighbors all share one color. Star moves are local, so they conserve the windings $W_i$. They are the elementary moves of our $\cramped{T=0}$ Monte Carlo [\cref{sm:mc}].

\begin{figure}[t]
  \includegraphics{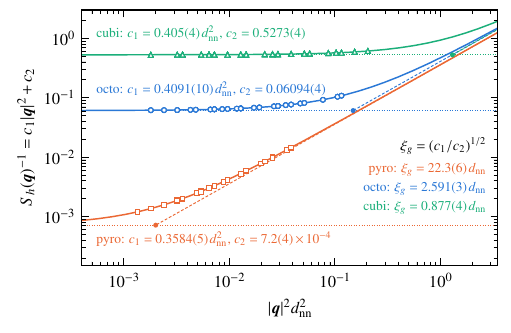}
  \caption{Inverse height structure factor of the $T=0$ ensembles for all system sizes and both initial states used in \cref{tab:heightfits}. Solid lines: straight-line fits $S_h^{-1}=c_1|\bm q|^2+c_2$ in $|\bm q|^2$, shown on logarithmic axes; dashed: $c_1|\bm q|^2$; dotted: $c_2$; dots: $|\bm q|=1/\xig$, where the two parts are equal.}
  \label{fig:heightSq}
\end{figure}

\paragraph{Anisotropy potential fits}
The stiffness $K$ and the locking strength $|g|$ of \cref{eq:free_energy} can be extracted from the fluctuations of the height field.
We measure the height structure factor,
\begin{equation}
    S_h(\bm q)=\langle|h(\bm q)|^2\rangle
\end{equation}
with the Fourier transformed height field defined as
\begin{equation}
    h(\bm q)=N_v^{-1/2}\sum_v h_v\,e^{-i\bm q\cdot\bm x_v}.
\end{equation}
At small $\bm q$, $\cramped{\cramped{S_h(\bm q)^{-1}=c_1|\bm q|^2+c_2}}$: the stiffness sets the slope $c_1$ and the locking sets the offset $c_2$.
Following Ref.~\cite{zengZeroTemperaturePhase1997}, we use the microscopic heights $h_v$. Their sublattice modulation ($h$ even on A, odd on B) only contributes at nonzero reciprocal-lattice vectors, so $h(\bm q)$ matches the coarse-grained height at small $\bm q$, and we fit this form for each size and start separately, at the smallest $|\bm q|$ (the same integers $n_i$ in $\cramped{\bm q=\sum_i n_i\bm b_i/L}$ at every system size $L$; see \cref{sm:Jq}).
\cref{tab:heightfits} quotes inverse-variance means over the sizes where the fits have converged: $L\geq48$ on the pyrochlore, $L\geq32$ on the octochlore and $L\geq24$ on the cubichlore lattice.
The pyrochlore $|g|$ error is widened to cover its remaining drift with $L$, and the cubichlore $|g|$ error is scaled by $\sqrt{\chi^2/\mathrm{dof}}$.
Here $\bm q$ is a physical wavevector, so $c_1$ has units of area and $c_2$ is dimensionless.
With the vertex density $n_v$, the stiffness and locking strength in units of $\dnn$ are
\begin{equation}
    K=c_1 n_v\dnn \quad \text{and} \quad |g|=c_2 n_v\dnn^3/4\pi^2
    \label{eq:Kg}
\end{equation}
and the locking length is $\xig=(c_1/c_2)^{1/2}$, or $\xig/\dnn=(K/4\pi^2|g|)^{1/2}$.
The fit gives only $|g|$; its sign follows from the angle at which $\varphi$ locks: $\langle\cos2\varphi\rangle=-0.993$ on the pyrochlore lattice and $+0.9999$ on the octochlore and cubichlore lattices.
\cref{fig:heightSq} shows the fits of $S_h^{-1}$, and \cref{tab:heightfits} lists the resulting $K$, $|g|$ and $\xig$.

\begin{table}[t]
  \caption{Stiffness $K$ and locking strength $|g|$ per volume [\cref{eq:free_energy}] in units of $\dnn$, locking length $\xig$, and sign of $g$ for the $T=0$ ensembles, obtained from the fitted $c_1$ and $c_2$ of \cref{fig:heightSq} via \cref{eq:Kg}.}
  \label{tab:heightfits}
  \begin{ruledtabular}
  \begin{tabular}{lcccc}
    & $K$ & $|g|$ & $\xig/\dnn$ & sign of $g$\\
    \hline
    pyrochlore & $0.2328(3)$ & $1.18(6)\times10^{-5}$ & $22.3(6)$ & $-$\\
    octochlore & $0.4091(10)$ & $1.5437(10)\times10^{-3}$ & $2.591(3)$ & $+$\\
    cubichlore & $0.526(5)$ & $1.7352(12)\times10^{-2}$ & $0.877(4)$ & $+$
  \end{tabular}
  \end{ruledtabular}
\end{table}

\paragraph{Extracting the locking angle} 
BSS order locks $\varphi$ to $0$ or $\pi$ and PSS order to $\pm\pi/2$, and the same angles appear in the $(\BB,\PP)$ histograms of \cref{fig:T0}(b).
For each $\cramped{T=0}$ sample we form the histogram angle $\theta_{\BB\PP}=\arg(\BB+i\kappa\PP)$, with $\cramped{\kappa=2.72}$ the relative scale of $\PP$ and $\BB$ used in \cref{fig:T0}(b), and compare $\langle\cos(2\theta_{\BB\PP})\rangle$ with $\langle\cos(2\varphi)\rangle$ measured from the coarse-grained heights.
The two agree closely on all three lattices and at all sizes (\cref{tab:heights}), and $\BB$ and $\PP$ are therefore good proxies for the height angle.
The small differences on the pyrochlore lattice at $L\le32$ arise because the histogram there is still a ring, whose angle depends on the choice of $\kappa$.

\begin{table}[h]
  \caption{$\langle\cos(2\theta_{\BB\PP})\rangle$ from the raw $(\BB,\PP)$ histograms ($\kappa=2.72$) and $\langle\cos(2\varphi)\rangle$ from the heights, at $\cramped{T=0}$.}
  \label{tab:heights}
  \begin{ruledtabular}
  \begin{tabular}{lcccc}
    & $L=16$ & $32$ & $64$ & $128$ \\
    \hline
    pyro, histogram & $-0.104$ & $-0.556$ & $-0.943$ & $-0.993$ \\
    pyro, heights   & $-0.139$ & $-0.581$ & $-0.948$ & $-0.993$ \\
    octo, histogram & $0.955$ & $0.994$ & $0.999$ & $1.000$ \\
    octo, heights   & $0.961$ & $0.995$ & $0.999$ & $1.000$ \\
    cubi, histogram & $0.991$ & $0.999$ & $1.000$ & $1.000$ \\
    cubi, heights   & $0.995$ & $0.999$ & $1.000$ & $1.000$ \\
  \end{tabular}
  \end{ruledtabular}
\end{table}

\begin{figure}[t]
  \setlength{\unitlength}{1pt}
  \begin{picture}(246,200)
    \put(0,0){\includegraphics{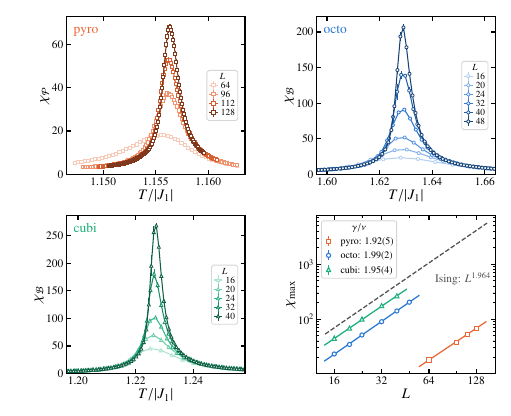}}
    \put(2,190){(a)}\put(124,190){(b)}\put(2,94){(c)}\put(124,94){(d)}
  \end{picture}
  \caption{Order-parameter susceptibilities near $T_c$. (a--c) $\chi_X=L^3(\langle X^2\rangle-\langle|X|\rangle^2)/T$ of $X=\PP$ on the pyrochlore ($L=64$--$128$) and $X=\BB$ on the octochlore and cubichlore lattices ($L\geq16$), on the simulated temperature grid (no reweighting). (d) Peak heights $\chi_{\max}$ (quadratic fit through the five points around the maximum) against $L$; lines: power-law fits, whose slopes $\gamma/\nu$ (legend; jackknife errors) agree with the 3D Ising value $\gamma/\nu=1.964$ (dashed)~\cite{hasenbuschFiniteSizeScaling2010}.}
  \label{fig:chi}
\end{figure}

\smsection{Finite-size scaling at the transition}{sm:fss}
This section explains how the numbers in \cref{tab:exponents} are obtained from the finite-temperature data.
Each estimate uses the order parameter $X$ that orders on a given lattice: $X=\BB$ on the octochlore and cubichlore lattices and $X=\PP$ on the pyrochlore lattice.
The critical temperature $T_c$ and the Binder crossing $U^*$ are defined by the points where the curves $U_X(T)$ of consecutive sizes cross, using $L\geq 16$ for the octochlore and cubichlore lattices and $L=64$--$128$ for the pyrochlore lattice.
$T_c$ and $U^*$ are the means of these consecutive-pair crossings.
The exponents $\gamma/\nu$, $\beta/\nu$ and $1/\nu$ follow from how the susceptibility peak, the order parameter at that peak, and the maximum slope of the Binder cumulant all scale with $L$~\cite{binderFiniteSizeScaling1981,ferrenbergPushingLimitsMonte2018}.
Here $\cramped{\chi_X=L^3(\langle X^2\rangle-\langle|X|\rangle^2)/T}$ [\cref{fig:chi}], the order parameter is $\langle|X|\rangle$ evaluated at the temperature of the $\chi_X$ peak, and $|\mathrm dU_X/\mathrm dT|$ is taken by finite differences on the simulated grid.
None of these requires $T_c$ as an input, which matters because quantities evaluated at a fixed temperature are very sensitive to the exact $T_c$ at these sizes.
All errors are jackknife estimates, as described in \cref{sm:mc}, except for $T_c$ on the octochlore and cubichlore lattices, where the error is the spread of the pair crossings.

\begin{figure}[t]
  \includegraphics{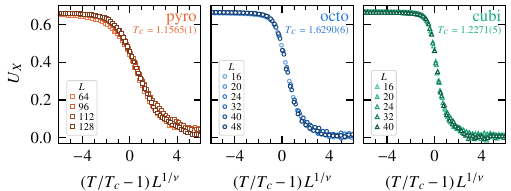}
  \caption{Scaling collapse of $U_X$ against $(T/T_c-1)L^{1/\nu}$ with the 3D Ising $\nu=0.630$, on the pyrochlore ($L\geq64$), octochlore and cubichlore ($L\geq16$) lattices; each lattice collapses onto its own curve, the widths differing by non-universal metric factors.}
  \label{fig:collapse}
\end{figure}

\paragraph{Data collapse}
A scaling collapse of $U_\BB$ gives a second estimate of $\nu$, from the same simulations.
We plot $U_\BB$ against $(T/T_c-1)L^{1/\nu}$ for $L\geq16$ and let $\nu$ and $T_c$ vary so that the curves of all sizes fall on top of each other, minimizing their weighted spread over $0.15<U_\BB<0.6$.
To compare sizes at the same temperature, we interpolate the data of each seed by multiple-histogram reweighting~\cite{ferrenbergOptimizedMonteCarlo1989}.
The collapse gives $\nu=0.625(10)$ on the octochlore and $0.608(14)$ on the cubichlore lattice, i.e.\ $1/\nu=1.60(3)$ and $1.64(4)$, close to the 3D Ising value $1.587$ and well above the 3D XY value $1.488$.
\cref{fig:collapse} shows the collapse with $\nu$ fixed to the Ising value $0.630$.

\paragraph{Other exponents}
The 3D Ising and XY values of $\gamma/\nu$ and $\beta/\nu$ differ by less than $0.2\%$, so these exponents cannot easily distinguish these two classes.
Our values agree with both, and the hyperscaling relation $2\beta/\nu+\gamma/\nu=d$ (last column of \cref{tab:exponents}) holds to within $1\%$.
The two classes are instead distinguished by $\nu$ (above) and by the sign of $\alpha$, positive for 3D Ising and negative for 3D XY, with the Binder crossings $U^*$ (below) as a consistency check.
Our specific-heat maxima grow with $L$, with $\alpha/\nu\approx0.2$--$0.3$ (3D Ising: $0.175$).
Because $\alpha$ is small, the singular part of the specific heat is hard to separate from its nonsingular background at these sizes, so we rely only on the sign of $\alpha$.

\paragraph{Binder crossings}
The Binder crossings (\cref{tab:exponents}) are compared with the 3D Ising value $U^*=0.4655$ for a periodic cubic box~\cite{hasenbuschFiniteSizeScaling2010}, for both the cubic cubichlore cell and the rhombohedral pyrochlore and octochlore cells.
The 3D XY value in \cref{tab:exponents}, $0.3785$, is that of one Cartesian component of an isotropic XY order parameter.
Since $U^*$ depends on the shape of the simulation cell~\cite{sorensenIsingModelPenrose1991}, we use it only as a consistency check: all three crossings lie close to the Ising value and well above the XY value (\cref{tab:exponents}).

\paragraph{Pyrochlore crossover}
This paragraph collects the numbers used to characterize 
the pyrochlore crossover of \cref{app:critical}.
The anisotropy $\cos(2\varphi)$ has RG eigenvalue $Y_2=1.7639(11)$ at the 3D XY fixed point~\cite{hasenbuschAnisotropicPerturbationsThreedimensional2011}.
At that fixed point, the O(2) moment ratio $\cramped{R_4\equiv\langle|\bm M|^4\rangle/\langle|\bm M|^2\rangle^2}$ takes the value $\cramped{R_4^*\approx1.243}$ in a cubic periodic box~\cite{campostriniCriticalBehaviorThreedimensional2001}.
For an isotropic two-component vector, this gives the Binder cumulant of one component as $\cramped{U^*=1-R_4^*/2\approx0.3785}$.
Beyond the meeting point at $L\approx18$, $U_\PP(T_c)$ rises to $0.44$--$0.45$ for $L=96$--$128$ [\cref{fig:finiteT}(d)].
Using only the largest pairs ($L\geq96$) moves the crossing from $0.44(1)$ to $0.46(3)$, closer to the Ising value.
The specific heat peak drifts accordingly: for $L=8$ it lies $0.02$ above $T_c$, crosses below $T_c$ between $L=16$ and $24$, and only then approaches $T_c$ from below for $\cramped{L>24}$ [\cref{fig:T0}(f)].
It is also worth commenting that the $q=3$ Potts antiferromagnet on the diamond lattice exhibits severe finite-size effects~\cite{lapinskasTwoTypesOrdering1998}.

\smsection{Numerical Methods and Entropy Estimation}{sm:mc}
\label{app:methods}
This section describes how the data in the main text were generated: Monte Carlo simulations of the spin model at finite and at zero temperature, and of the three-state Potts model for the entropies.

\paragraph{Finite temperature}
All simulations use the Ising Hamiltonian of \cref{eq:H} with the couplings of \cref{tab:couplings} on periodic $L^3$ boxes.
At finite $T$ we use replica-exchange Monte Carlo~\cite{swendsenReplicaMonteCarlo1986,hukushimaExchangeMonteCarlo1996} with two kinds of updates: single-spin Metropolis flips, and star moves, which flip a fully polarized cluster [\cref{sm:heights_mc}].
The simulations run on the multispin-coded~\cite{jacobsMultispinCodingVery1981} GPU Ising codes developed in~\cite{burkeSpinVorticityModels}.
For each lattice and size we ran four independent seeds from random starts, on a linear temperature grid around $T_c$ (\cref{tab:mc}).
Errors come from a leave-one-seed-out jackknife~\cite{efronJackknifeBootstrapOther1982,youngEverythingYouWanted2015}.
For the largest pyrochlore sizes, $L=64$--$128$, the data are instead generated from one long run per size on a narrower grid around $T_c$, and their errors from a jackknife over 20 consecutive blocks of the time series.

\paragraph{Zero temperature}
At $T=0$ we sample the ground states with star moves alone.
These moves never leave the ground-state manifold and conserve the windings $W_i$ [\cref{sm:heights_potts}]; all runs start with $W_i=0$ and so remain in that sector.
For each lattice we simulate $L=16$ to $128$ from two starts, PSS- and BSS-polarized, with 32 independent runs each.
The two starts give the same results within errors, which shows that the runs have equilibrated.
The runs last $2\times10^5$ sweeps on the octochlore and cubichlore lattices and up to $1.2\times10^6$ sweeps on the pyrochlore lattice.
We discard the first half of each run, and errors are the standard error over the 32 runs.

\paragraph{Residual entropy}
We obtain the SVM entropies $S_0$ of \cref{tab:entropy} by thermodynamic integration, starting from the entropy $\ln2$ per spin at infinite temperature.
With $E$ the energy per spin, measured from the ground state, the entropy per spin at $T=0$ is
\begin{equation}
S_0=\ln2-\tfrac12k_2\beta_0^2+\big[\beta E\big]_{\beta_0}^{\beta_1}-\int_{\beta_0}^{\beta_1}E\,\mathrm d\beta .
\label{eq:TI}
\end{equation}
The first two terms are the high-temperature expansion of the entropy at $\beta_0=1/300$, with $k_2=\tfrac12\sum nJ^2$ summed over the shells of \cref{tab:couplings} ($n$ neighbors with coupling $J$).
We integrate up to $\beta_1=5$ ($T=0.2$) with a cubic spline through about $200$ temperatures, uniform in $\beta$ and refined near $T_c$, from replica-exchange runs at $L=32$ and $48$ ($L=32$ and $40$ on the cubichlore lattice).
The entropy left below $T=0.2$ is at most about $2\times10^{-6}$, within our errors.
The table quotes the largest size; its errors include the spread over seeds, the integration rule, the choice of $\beta_0$, and the difference between sizes.
For all three lattices, the result agrees within uncertainty with the Potts value $2s_v/z$.

\paragraph{Colorings and Potts entropies}
The Potts entropies count proper colorings, whereas the SVM ground states are height fields; here we explain when the two counts agree.
To turn a proper coloring back into heights, set $h_b-h_a=\pm1$ according to whether $c_b-c_a=\pm1\bmod3$.
This is consistent only if the steps around every plaquette sum to zero, as required by $\omega_p=0$, whereas the coloring fixes each sum only modulo 3.
Four steps of $\pm1$ never sum to a nonzero multiple of 3, so on the square plaquettes of the simple-cubic and bcc lattices every proper coloring gives a ground state.
On these lattices the SVM ground-state entropy therefore follows directly from the ground-state entropy of the $\cramped{q=3}$ Potts antiferromagnet $H_{\textrm P}$ [\cref{app:potts}],
\begin{equation}
    s_v=\frac{\ln(\#\,\text{proper 3-colorings)}}{N_v},
\end{equation}
listed in \cref{tab:entropy}.
On a hexagon of the diamond lattice, however, the proper coloring $0,1,2,0,1,2$ gives $\omega_p=\pm6$; we call such hexagons chiral.
Colorings with chiral hexagons therefore do not map to SVM ground states, and the Potts ground-state count for the diamond lattice must be corrected for these.

\begin{table}[t]
  \caption{Finite-temperature runs: sizes, number of temperatures $N_T$, temperature window, thermalization sweeps and measurements (one every four sweeps) per seed; four seeds each.}
  \label{tab:mc}
  \begin{ruledtabular}
  \begin{tabular}{lccccc}
    & $L$ & $N_T$ & $T/|J_1|$ at $L_{\max}$ & thermalization & measurement \\
    \hline
    pyro & 8--48  & 33--169 & 1.06--1.25 & $2$--$5\times10^4$ & $2\times10^4$ \\
    octo & 8--48  & 36--84  & 1.57--1.67 & $2$--$4.4\times10^4$ & $2$--$2.8\times10^4$ \\
    cubi & 8--40  & 35--81  & 1.19--1.26 & $2$--$5.2\times10^4$ & $2$--$3.2\times10^4$ \\
  \end{tabular}
  \end{ruledtabular}
\end{table}

To do so we weight each proper coloring by a fugacity $y$ per chiral hexagon,
\begin{equation}
Z(y)=\sum_{\text{proper colorings}}y^{N_6},
\end{equation}
where $N_6$ is the number of chiral hexagons in a coloring.
$\cramped{Z(y=1)}$ counts all proper colorings, and $\cramped{Z(y=0)}$ counts only those without chiral hexagons.
Since $\mathrm d(\ln Z)/\mathrm dy=\langle N_6\rangle_y/y$, a thermodynamic integration in $y$, analogous to \cref{eq:TI}, gives the entropy per vertex removed by the constraint,
\begin{equation}
\Delta_y=\frac1{N_v}\int_0^1\frac{\langle N_6\rangle_y}{y}\,\mathrm dy.
\end{equation}
On the diamond lattice we find $\Delta_y=6.42(2)\times10^{-4}$.
Subtracting it from the entropy of all proper colorings, $0.41142(2)$, gives $s_v=0.41078(2)$ for the colorings without chiral hexagons (\cref{tab:entropy}).
This small correction reduces the difference between the Potts prediction $2s_v/z$ and the thermodynamic-integration value $S_0$ on the pyrochlore lattice by a factor of about 15, from $3\times10^{-4}$ to $2\times10^{-5}$. We next describe the simulations behind these values.

The entropy of all proper colorings comes from simulating the Potts model at nonzero $T$ and counting its states at zero energy.
The simulations combine heat-bath updates with the Wang--Swendsen--Koteck\'y (WSK) cluster algorithm~\cite{wangAntiferromagneticPottsModels1989a,wangThreestateAntiferromagneticPotts1990}.
From the data, we reconstruct the density of states $g(E)$ by multiple-histogram reweighting~\cite{ferrenbergOptimizedMonteCarlo1989}, normalize it to the $3^{N_v}$ states at infinite temperature, and extract $\ln g(0)$.
We simulate sizes $L=4$--$12$ and extrapolate the entropy per vertex to infinite size, assuming finite-size corrections proportional to $1/N_v$.
Our values of $s_v$ for all proper colorings agree with earlier Potts estimates~\cite{yamaguchiThreedimensionalAntiferromagneticQstate2001a,lapinskasTwoTypesOrdering1998}.

For the chiral-hexagon correction on the diamond lattice we simulate the Potts model at $T=0$, so that only proper colorings appear, and sample them at fixed $y$ with Kempe-chain moves~\cite{moharNewKempeInvariant2009}.
A Kempe chain is a connected cluster of sites carrying two of the three colors; swapping those two colors gives another proper coloring, which we accept with the Metropolis probability, $\min(1,y^{\Delta N_6})$, where $\Delta N_6$ is the change in the number of chiral hexagons.
We measure $\langle N_6\rangle_y$ at twelve values of $y$ between $0.1$ and $1$, more densely near $y=1$ where the integrand is largest, at $L=24$ and $32$.
At small $y$ we find $\langle N_6\rangle_y\sim y^6$, so the integrand $\langle N_6\rangle_y/y$ vanishes at $y=0$.
We add $y=0$ as the first node and integrate with Simpson's rule; the interval $0\le y<0.1$ contributes less than $10^{-9}$ to $\Delta_y$.
The diamond $s_v$ in \cref{tab:entropy} is the restricted value obtained with this $\Delta_y$ for direct comparison with the SVM entropies.

\end{document}